\documentclass[a4paper,12pt]{article}
\usepackage[cp1251]{inputenc}

\usepackage{amsfonts,amssymb,amsthm,mathtools}
\usepackage{amsmath}
\usepackage{icomma}
\usepackage{latexsym}
\usepackage{graphicx}
\usepackage{subfigure}
\usepackage {bm}

\usepackage{geometry} 
\begin{document}

\begin{center}
\textbf{QUANTUM MECHANICS OF  STATIONARY STATES OF PARTICLES IN A SPACE-TIME OF CLASSICAL BLACK HOLES}
\end{center}

\begin{center}
{M.~V.~Gorbatenko$^{1}$, V.~P.~Neznamov$^{1,2}$\footnote{vpneznamov@mail.ru, vpneznamov@vniief.ru}}
\end{center}

\begin{center}
$^{1}$FSUE ''RFNC-VNIIEF'', Russia, Sarov, Mira pr., 37, 607188 \\
$^{2}$National Research Nuclear University "MEPHI", Moscow, Russia
\end{center}


\begin{abstract}

We consider interactions of scalar particles, photons, and 
fermions in Schwarzschild, Reissner-Nordstr\"{o}m, 
Kerr, and Kerr-Newman gravitational and electromagnetic fields with a zero and nonzero cosmological constant. We also consider interactions of scalar particles, photons, and fermions with nonextremal 
rotating charged black holes in a minimal five-dimensional gauge supergravity. We analyze the behavior of effective potentials in second-order relativistic Schr\"{o}dinger-type equations. In all cases, 
we establish the existence of the regime of particle "falling" on event horizons. An alternative can be collapsars with fermions in stationary bound states without a regime of particles "falling".\\

DOI: 10.1134/S0040577920110070

\end{abstract}

\small{\textbf{Keywords:} quantum-mechanical hypothesis of cosmic censorship, Schr\"{o}dinger-type equation, effective potential, scalar particle, photon, fermion, Schwarzschild, Reissner-Nordstr\"{o}m, Kerr, and Kerr-Newman black holes with zero and nonzero cosmological constant, anti-de Sitter black hole in five-dimensional supergravity}

\section{Introduction}

For a closed system of ''a particle in an external force field'', quantum 
mechanics admits the existence of stationary states with certain particle 
energies. Stationary states include both states of a discrete spectrum 
(bound states) and states of a continuous spectrum (scattering states).
In this case, the particle wave function is written in the form

\begin{equation}
\label{eq1}
\psi \left( {{\rm {\bf r}},t} \right)=\psi \left( {{\rm {\bf r}}} 
\right)e^{-i\,E\,t},
\end{equation}
where $E$ is real energy of particle. Here and hereafter, we use the system 
of units with $\hslash =c=1$.

Here, we consider interactions of scalar particles $\left( {S=0} 
\right)$, photons$\left( {S=1} \right)$, and fermions $\left( {S=1 
\mathord{\left/ {\vphantom {1 2}} \right. \kern-\nulldelimiterspace} 2} 
\right)$ with the Schwarzschild, Reissner-Nordstr\"{o}m, Kerr, and 
Kerr-Newman black holes with a zero and a nonzero cosmological constant. For 
the listed metrics, we separate the variables in the Klein-Gordon and 
Maxwell equations. We bring the equations for the radial functions to the 
form of second-order relativistic Schr\"{o}dinger-type equations with 
effective potentials. An analogous procedure was performed using a second-order self-adjoint equation with a spinor wave function for fermions \cite{1}. In addition, we analyze the behavior of effective potentials in neighborhoods of event horizons.
We similarly analyze interactions of scalar particles, photons, and fermions with 
nonextremal rotating charged black holes in a minimal five-dimensional 
gauge supergravity.

The existence of a regime of particle ''falling'' \cite{2}, \cite{3} on event horizons was 
established for all considered metrics and for particles with different spins. Separate states of the considered particles with the energy $E^{ext}$ for extremal black holes and degenerate bound states with $E^{st}$ for fermions are an exception \cite{4} - \cite{6}.

For the Schwarzschild, Painlev\'{e} -Gullstrand and Kerr metrics, representation (\ref{eq1}) was previously used in many papers to prove the existence of 
nonstationary solutions of the Klein-Gordon and Dirac equations 
corresponding to bound states of spin and spinless particles with complex 
energies decaying in time (see, e.g., \cite{7} - \cite{17}).
On the other hand,  the absence of physically meaningful stationary 
solutions (with real energy) of the Dirac equation in classical 
Schwarzschild, Reissner-Nordstr\"{o}m, Kerr, and Kerr-Newman fields was 
proved in \cite{18} - \cite{21}.

The presented results are easily explained by the existence of a regime of particles "falling" on event horizons for all classical black holes, which we proved. As an alternative, the existence of composite systems, collapsars with fermions in degenerate stationary bound states, is possible \cite{4} - \cite{6}.

This paper is organized as follows. In Sec. 2, for convenience of analysis, we supplement the quantum mechanical hypothesis of cosmic censorship, previously practically introduced in \cite{22}, with numerical characteristics. In more detail, we reveal the content of the regime of a particle ''falling'' on a singular center, unacceptable for quantum theory. We show that quantum 
mechanical hypothesis of cosmic censorship holds in the example of the problem ''Z\textgreater 137 catastrophe'' in hydrogen-like atoms \cite{23}. In Secs. 3 and 4, we study the interaction of scalar particles with Schwarzschild, Painlev\'{e}-Gullstrand, Reissner-Nordstr\"{o}m, Kerr, and 
Kerr-Newman black holes with a zero and a nonzero cosmological constant. In 
Sec. 5, we study this problem for five-dimensional anti-de Sitter black holes. In 
Secs. 6 and 7, we discuss our results.

We choose the metric signature of the Minkowsky space-time equal to \,\,\,\,\,\,\,$\eta_{\alpha \beta } =\mbox{diag}\left[ {1,-1,-1,-1} \right]$.


\section{Quantum mechanical hypothesis of cosmic censorship }
\label{sec:2}

In classical physics, the hypothesis of cosmic censorship, proposed by 
Penrose \cite{24}, forbids the existence  in Nature of singularities not covered by event horizons.
A quantum mechanical hypothesis of cosmic censorship was practically proposed in \cite{22}, in the introduction of which the authors wrote, ''... we will say that a system is nonsingular when the evolution of any state is uniquely defined for all time. If this is not the case, then there is some loss of 
predictability and we will say that the system is singular''. By analogy with Penrose \cite{24}, we must add that such singular systems cannot exist in Nature.

We present some numerical characteristics of singular and nonsingular 
systems. For second-order radial equations brought to the form of 
Schr\"{o}dinger-type equations with effective potentials $U_{eff} \left( 
\rho \right)$, the behavior of these potentials in neighborhoods of event 
horizons is important. For all considered metrics, the behavior of effective potentials in neighborhoods of event horizons often has the form of an infinitely deep potential well:

\begin{equation}
\label{eq3}
\left. {U_{eff} \left( \rho \right)} \right|_{\rho \to \rho_{\pm } } 
=-\,\frac{K_{1} }{\left( {\rho -\rho_{\pm } } \right)^{2}}.
\end{equation}
If $K_{1/8}$, then the so-called mode of particle ''falling'' on the 
event horizon occurs \cite{2} - \cite{6}. In this case, the system is singular. The radial function of the Schr\"{o}dinger-type equation behaves as

\begin{equation}
\label{eq4}
\left. {R\left( \rho \right)} \right|_{\rho \to \rho_{\pm } } \sim \left( 
{\rho -\rho_{\pm } } \right)^{1/2}\sin \left( {\sqrt {K_{2} } \log \left( {\rho 
-\rho_{\pm } } \right)+\delta } \right),
\end{equation}
where $K_{2} =2\left( {K_{1} -\left( {1/8} \right)} \right)$.
As $\rho \to \rho_{\pm } $, the radial functions $R\left( \rho \right)$ of stationary states of 
the discrete and continuous spectra have an infinite number of zeros, and discrete energy levels appear and ''dive'' beyond the permitted domains of functions $R\left( \rho \right)$. At  $\rho 
=\rho_{\pm } $, the functions $R\left( \rho \right)$ have no definite values.

The system is also singular if the exponent of the denominator in (\ref{eq3}) exceeds two. In this case,

\begin{equation}
\label{eq5}
\left. {R_{s} \left( \rho \right)} \right|_{\rho \to \rho_{\pm } } \sim 
\left( {\rho -\rho_{\pm } } \right)^{s \mathord{\left/ {\vphantom {s 4}} 
\right. \kern-\nulldelimiterspace} 4}\sin \left( {\frac{2}{s-2}\sqrt 
{\frac{K_{1} }{\left( {\rho -\rho_{\pm } } \right)^{s-2}}} +\delta_{s} } 
\right).
\end{equation}
In formulas (\ref{eq4}) and (\ref{eq5}), $\delta$, and $\delta_{s} $ are arbitrary phases $\left( {0\le 
\delta ,\,\,\delta_{s} <\pi } \right)$ and $s>2$ is the exponent in the 
expression for the effective potential $\left. {U_{eff} \left( \rho 
\right)} \right|_{\rho \to \rho_{\pm } } =-\,\,{K_{1} } \mathord{\left/ 
{\vphantom {{K_{1} } {\left( {\rho -\rho_{\pm } } \right)^{s}}}} \right. 
\kern-\nulldelimiterspace} {\left( {\rho -\rho_{\pm } } \right)^{s}}$.

If $K_{1} <1/8$ and $s=2$, then the system is nonsingular. In this 
case, the existence of stationary bound states of particles with 
$\varepsilon <1_{\, }$ is possible.

In the Hamiltonian formalism, the mode of particle ''falling'' on an  event 
horizon corresponds to the fact that the Hamiltonian $H$ has nonzero deficiency indexes 
\cite{25} - \cite{27}. To eliminate of this mode, we must choose additional boundary conditions on event horizons. A self-adjoint extension of the Hermitian operator $H$ is defined by this choice.

In the history of quantum mechanics, there is an example confirming the quantum 
mechanical hypothesis of cosmic censorship. For hydrogen-like atoms, the 
Sommerfeld formula for the fine structure of energy levels has the form

\begin{equation}
\label{eq6}
\varepsilon =\left( {1+\frac{\alpha_{fs }^{2} Z^{2}}{\left( {n-\left| 
\kappa \right|+\sqrt {\kappa^{2}-\alpha_{fs }^{2} Z^{2}} } \right)}} 
\right)^{-\,1 \mathord{\left/ {\vphantom {1 2}} \right. 
\kern-\nulldelimiterspace} 2},
\end{equation}
where $\alpha_{fs }$ is the electromagnetic fine structure constant, $n$ is the principal quantum number, $\kappa $ is the quantum number of the Dirac equation,

\[\kappa =\mp 1,\mp 2,...=\left\{ {\begin{array}{l}
 -\left( {l+1} \right),\,\,\,\,j=l+\dfrac{1}{2}, \\ 
 \,\,\,\,\,\,\,\,\,l,\,\,\,\,\,\,\,\,\,j=l-\dfrac{1}{2}, \\ 
 \end{array}} \right.,
\]
and $j$ and $l$ are the quantum numbers of the total and orbital angular momentum of a spin -1/2
particle. For $Z>137\left| \kappa \right|$, expression (\ref{eq6}) becomes complex 
(''$Z>137$ catastrophe'').

We consider solutions of a Schr\"{o}dinger-type equation with an effective potential for fermions in a Coulomb field \cite{28}.  The asymptotic formula for the effective potential as $\rho \to 0$ has the form

\begin{equation}
\label{eq7}
\left. {U_{eff}^{C} } \right|_{\rho \to 0} =-\frac{\left( {Z\alpha_{fs 
} } \right)^{2}-(3/4)+\left( {1-\kappa^{2}} \right)}{2\rho^{2}}.
\end{equation}
We can distinguish three typical domains depending on Z in asymptotic formula (\ref{eq7}). As an example, we consider these domains for the bound states $1S_{1/2} \left( {\kappa =-1} \right)$ and $2P_{1/2} \left( {\kappa =+1} \right)$. In the first domain $1\le Z<{\sqrt 3 } 
\mathord{\left/ {\vphantom {{\sqrt 3 } {2\alpha_{fs} }}} \right. 
\kern-\nulldelimiterspace} {2\alpha_{fs} }_{\, }$, there 
exists a positive barrier $\sim 1 \mathord{\left/ {\vphantom {1 {\rho 
^{2}}}} \right. \kern-\nulldelimiterspace} {\rho^{2}}_{\, }$ followed by potential well as $\rho \to 0$. The potential barrier disappears at $Z=Z_{cr} ={\sqrt 3 } \mathord{\left/ 
{\vphantom {{\sqrt 3 } {2\alpha_{fs} }}} \right. \kern-\nulldelimiterspace} 
{2\alpha_{fs} }\approx 118.7$, and the potential well $-K 
\mathord{\left/ {\vphantom {K {\rho^{2}}}} \right. 
\kern-\nulldelimiterspace} {\rho^{2}}_{\, }$ persists for $Z>Z_{cr}$ as $\rho \to 0$. In the second domain
$119\le Z<137$, we have the coefficient $K<1/8$, which admits the existence of fermionic stationary bound states \cite{2}, \cite{3}. In the third domain $Z\ge 137_{\, }$, there exists a potential well with 
$K\ge 1/8$ as $\rho \to 0$, which indicates the realization of a regime of ''falling'' on the 
center \cite{2}, \cite{3}. We show the dependencies $U_{eff}^{F} \left( \rho 
\right)$ for $\kappa =-1\left( {1S_{1 \mathord{\left/ {\vphantom {1 2}} \right. \kern-\nulldelimiterspace} 2} } \right)$ in Fig. 1 for $Z=1,\,119,\,\,140$. We also show the dependencies of the Coulomb potential $V\left( \rho \right)=-\left( {{Z\alpha_{fs} } \mathord{\left/ {\vphantom {{Z\alpha_{fs} 
} \rho }} \right. \kern-\nulldelimiterspace} \rho } \right)$ for comparison.

In the third domain with $\left( {Z\alpha_{fs } } \right)^{2}\ge \kappa 
^{2}$, the system ''fermion in a Coulomb field'' is singular. To eliminate the mode of ''falling'' on the center, it was proposed to take the finite dimensions of an atomic nucleus into account \cite{29} - 
\cite{32}. As a result, a cutoff of either the Coulomb or the effective potential occurs at characteristic lengths of the nuclei size (see Fig.1b). There are now about 30 such cutoff methods (see, e.g., \cite{33}).

The system ''an electron in a Coulomb field of an atomic nucleus of 
finite size'' is nonsingular.

\begin{figure}[h!]
\begin{minipage}[h]{0.5\linewidth}
\center{\includegraphics[width=1\linewidth]{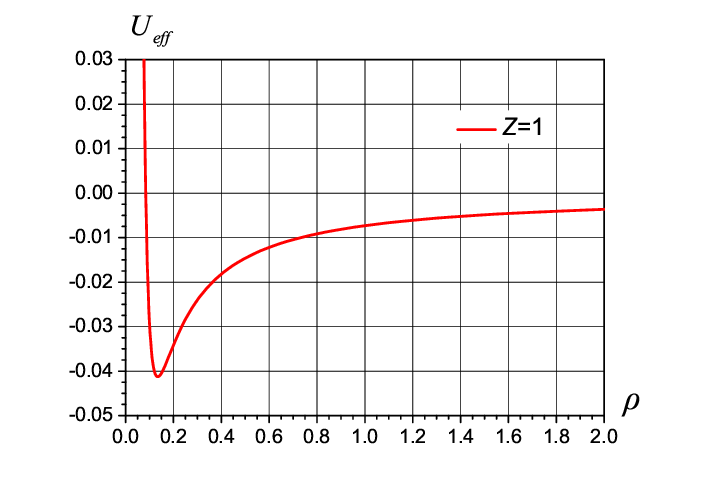}} \\ a)
\end{minipage}
\hfill
\begin{minipage}[h]{0.5\linewidth}
\center{\includegraphics[width=1\linewidth]{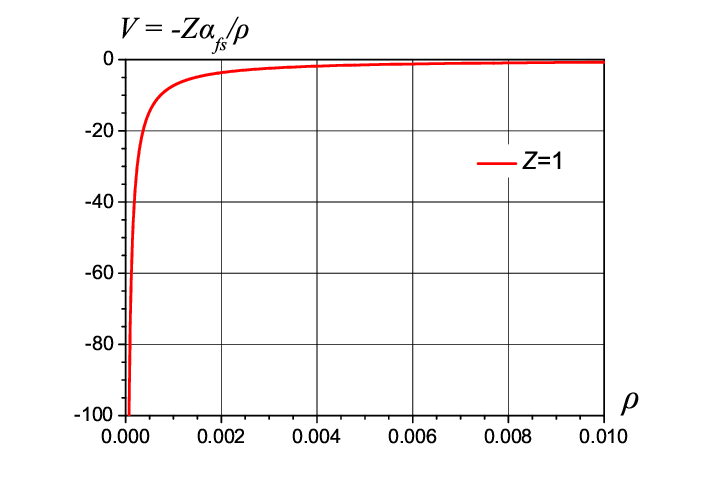}} \\b)
\end{minipage}
\vfill
\begin{minipage}[h]{0.5\linewidth}
\center{\includegraphics[width=1\linewidth]{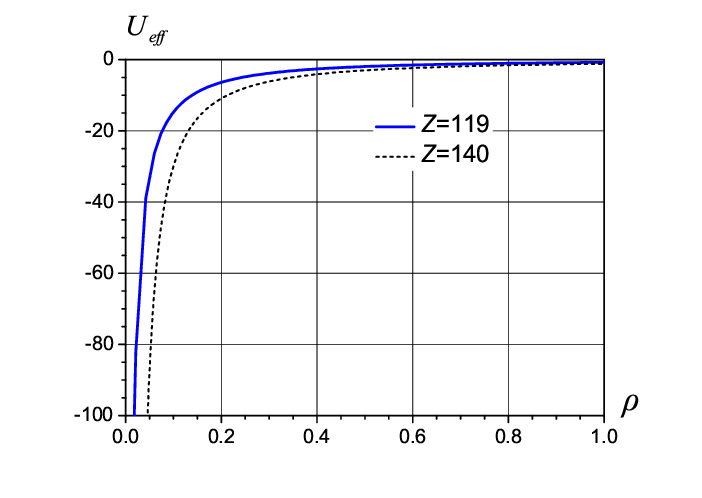}}  \\ c)
\end{minipage}
\hfill
\begin{minipage}[h]{0.5\linewidth}
\center{\includegraphics[width=1\linewidth]{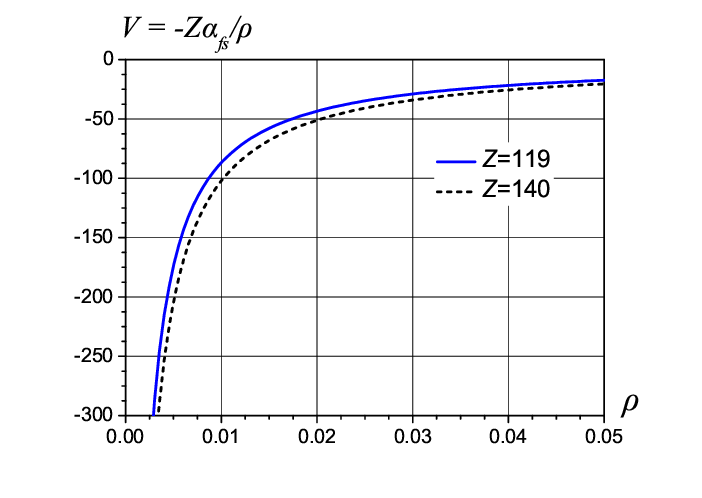}}  \\ d)
\end{minipage}
\caption{Dependences  $U_{eff} \left( \rho \right)$ and $V\left( \rho \right)$: a) $\left. U_{eff} \right|_{\rho\to 0 } =- \frac{\left( Z \alpha_{fS} \right)^2 - (3/4) + \left( 1 - \kappa^2\right)}{2 \rho^2}, \newline \kappa=-1 ,\,\,\,E/mc^2 \approx 1 $; c) $Z=Z_{cr} = \sqrt 3 / 2\alpha_{fs} \approx 118.7,\left. U_{eff} \right|_{\rho\to {0} } = - K/ \rho^2$.}
\label{fig:Figure1}
\end{figure}



\section{Metrics with a zero cosmological constant}
\label{sec:3}

\subsection{Brief characteristics of General Relativity solutions and the notation used in the paper}

{\bf{3.1.1} {\bf{Schwarzschild metric}}}. In the spherical coordinates $\left( {t,r,\theta ,\varphi } \right)_{\, }$, the Schwarzschild metric has the form

\begin{equation}
\label{eq8}
ds^{2}=f_{S} dt^{2}-\frac{dr^{2}}{f_{S} }-r^{2}\left( {d\theta^{2}+\sin 
^{2}\theta d\varphi^{2}} \right),
\end{equation}
where $f_{S} =1-\left( {{r_{0} } \mathord{\left/ {\vphantom {{r_{0} } r}} 
\right. \kern-\nulldelimiterspace} r} \right)$, $r_{0} ={2GM} 
\mathord{\left/ {\vphantom {{2GM} {c^{2}}}} \right. 
\kern-\nulldelimiterspace} {c^{2}}$ is the gravitation radius (event 
horizon), $G$ is the gravitation constant, $M$ is the mass of the gravitational field of a pointlike source, and $c$ is the speed of light.

{\bf{3.1.2} {\bf{Painlev\'{e}-Gullstrand metric}}}. We write the coordinates as $\left( {T,r,\theta ,\varphi } \right)$. The coordinate transformation of the Schwarzschild metric in spherical 
coordinates has the form

\begin{equation}
\label{eq10}
dT=dt-\sqrt {\frac{r_{0} }{r}} \frac{dr}{{1-r_{0} / r} }.
\end{equation}
The interval squared is defined as

\begin{equation}
\label{eq11}
ds^{2}=f_{S} dT^{2}-2\sqrt {\frac{r_{0} }{r}} dTdr-dr^{2}-r^{2}\left( 
{d\theta^{2}+\sin^{2}\theta d\varphi^{2}} \right).
\end{equation}

{\bf{3.1.3}} {\bf{Reissner-Nordstr\"{o}m metric}}. The static Reissner-Nordstr\"{o}m metric is characterized by a pointlike source with a mass $M$ and a charge $Q$:

\begin{equation}
\label{eq12}
ds^{2}=f_{RN} dt^{2}-\frac{dr^{2}}{f_{RN} }-r^{2}\left( {d\theta^{2}+\sin 
^{2}\theta d\varphi^{2}} \right),
\end{equation}
where $f_{RN} =1-r_{0} / r + r_{Q}^{2} / r^{2}$ and $r_{Q} = \sqrt G Q / c^{2}$.

If $r_{0}^{2} >4r_{Q}^{2} $, then

\begin{equation}
\label{eq13}
f_{RN} =\left( {1-\frac{\left( {r_{+} } \right)_{RN} }{r}} \right)\left( 
{1-\frac{\left( {r_{-} } \right)_{RN} }{r}} \right),
\end{equation}
where $\left( {r_{\pm } } \right)_{RN} $ are the radii of outer and inner event horizons,

\begin{equation}
\label{eq14}
\left( {r_{\pm } } \right)_{RN} =\frac{r_{0} }{2}\pm \sqrt {\frac{r_{0}^{2} 
}{4}-r_{Q}^{2} } .
\end{equation}

The case $r_{0}^{2} =4r_{Q}^{2} $ corresponds to an extremal Reissner-Nordstr\"{o}m field with a single event horizon $\left( {r_{+} } \right)_{RN} =\left( {r_{-} } \right)_{RN} 
={r_{0} } \mathord{\left/ {\vphantom {{r_{0} } 2}} \right. 
\kern-\nulldelimiterspace} 2$.

The case $r_{0}^{2} <4r_{Q}^{2} $ corresponds to a naked singularity. In this case, $f_{RN} >0$.

{\bf{3.1.4}} {\bf{Kerr and Kerr-Newman metrics}}. The stationary Kerr-Newman metric is characterized by a pointlike source with the mass $M$ and charge $Q$ rotating with the angular momentum ${\rm {\bf J}}=Mc{\rm {\bf a}}$. The Kerr metrics is the uncharged Kerr-Newman metric 
$\left( {Q=0} \right)$.

We can represent the Kerr-Newman metric in the Boyer-Lindquist coordinates $\left( 
{t,r,\theta ,\varphi } \right)_{\, }$  \cite{34} in the form

\begin{equation}
\label{eq15}
\begin{array}{c}
 ds^{2}=\left( {1-\dfrac{r_{0} r-r_{Q}^{2} }{r_{K}^{2} }} 
\right)dt^{2}+\dfrac{2a\left( {r_{0} r-r_{Q}^{2} } \right)}{r_{K}^{2} }\sin 
^{2}\theta dtd\varphi -\dfrac{r_{K}^{2} }{\Delta_{KN} }dr^{2}-r_{K}^{2} 
d\theta^{2}- \\ [10pt]
 -\left( {r^{2}+a^{2}+\dfrac{a^{2}\left( {r_{0} r-r_{Q}^{2} } 
\right)}{r_{K}^{2} }\sin^{2}\theta } \right)\sin^{2}\theta d\varphi^{2}, 
\\ 
 \end{array}
\end{equation}
where $r_{K}^{2} =r^{2}+a^{2}\cos^{2}\theta$ and $\Delta_{KN} 
=r^{2}f_{KN} =r^{2}\left( {1-r_{0} / r + \left( r_{Q}^{2} 
+a^{2} \right)  / r^{2}} \right)$.

If $r_{0} >2\sqrt {a^{2}+r_{Q}^{2} } $, then

\begin{equation}
\label{eq16}
f_{KN} =\left( {1-\frac{\left( {r_{+} } \right)_{KN} }{r}} \right)\left( 
{1-\frac{\left( {r_{-} } \right)_{KN} }{r}} \right),
\end{equation}
where $\left( {r_{\pm } } \right)_{KN} $ are the radii of outer and inner 
event horizons of the Kerr-Newman field,

\begin{equation}
\label{eq17}
\left( {r_{\pm } } \right)_{KN} =\frac{r_{0} }{2}\pm \sqrt {\frac{r_{0}^{2} 
}{4}-a^{2}-r_{Q}^{2} } .
\end{equation}

The case $r_{0} =2\sqrt {a^{2}+r_{Q}^{2} } ,\,\,\,\left( {r_{+} } 
\right)_{KN} =\left( {r_{-} } \right)_{KN} ={r_{0} } \mathord{\left/ 
{\vphantom {{r_{0} } 2}} \right. \kern-\nulldelimiterspace} 2$ corresponds 
to an extremal Kerr-Newman field.

The case $r_{0} <2\sqrt {a^{2}+r_{Q}^{2} } $ corresponds to a naked 
singularity of the Kerr-Newman field. In this case, $f_{KN} >0$.

For $Q=0$ , the Kerr-Newman metrics becomes the Kerr metric with

\begin{equation}
\label{eq18}
\Delta_{K} =r^{2}f_{K} =r^{2}\left( {1-\frac{r_{0} 
}{r}+\frac{a^{2}}{r^{2}}} \right).
\end{equation}

If $r_{0}^{2} >4a^{2}$, then

\begin{equation}
\label{eq19}
f_{K} =\left( {1-\frac{\left( {r_{+} } \right)_{K} }{r}} \right)\left( 
{1-\frac{\left( {r_{-} } \right)_{K} }{r}} \right),
\end{equation}
where $\left( {r_{\pm } } \right)_{K} =r_{0} / 2 \pm \sqrt {r_{0}^{2} 
/ 4-a^{2}}$.

For $r_{0} =2a$, we have $\left( {r_{+} } \right)_{K} =\left( {r_{-} } 
\right)_{K} ={r_{0} } \mathord{\left/ {\vphantom {{r_{0} } 2}} \right. 
\kern-\nulldelimiterspace} 2$. This case corresponds to an exremal Kerr field.

The case $r_{0} <2a$ corresponds to a naked singularity of the Kerr 
field. In this case, $f_{K} >0$.

It what follows, we write second-order equations for particles with the energy $E$, mass $m$, and electric charge $q$ in space-time of metrics (\ref{eq8}), (\ref{eq11}), (\ref{eq12}), and (\ref{eq15}) in dimensionless variables 

\begin{equation}
\label{eq21}
\begin{array}{c}
 \rho =\dfrac{r}{l_{c} },\,\,\,\varepsilon =\dfrac{E}{mc^{2}},\,\,\,\alpha 
=\dfrac{r_{0} }{2l_{c} }=\dfrac{GMm}{\hslash c}=\dfrac{Mm}{M_{P}^{2} 
},\\  [10pt]
\alpha_{Q} =\dfrac{r_{Q} }{l_{c} }=\dfrac{\sqrt G Qm}{\hslash 
c}=\dfrac{\sqrt {\alpha_{fs} } }{M_{P} }m\dfrac{Q}{e}, 
 \,\,\,\alpha_{a} =\dfrac{a}{l_{c} },\,\,\,\alpha_{em} =\dfrac{qQ}{\hslash 
c}=\alpha_{fs} \dfrac{qQ}{e^{2}}, \\ 
 \end{array}
\end{equation}
where $l_{c} =\hslash \mathord{\left/ {\vphantom {\hslash {mc}}} \right. 
\kern-\nulldelimiterspace} {mc}$ is the particle Compton wavelength, $_{\, }M_{P} =\sqrt {{\hslash c} \mathord{\left/ {\vphantom 
{{\hslash c} G}} \right. \kern-\nulldelimiterspace} G} =2.2\cdot 
10^{-5}\,\,\mbox{g}$ \,\,\,$\left( {1.2\cdot 10^{19}\mbox{GeV}} \right)$ is the 
Planck mass, $\alpha_{fs} ={e^{2}} \mathord{\left/ {\vphantom {{e^{2}} 
{\hslash c}}} \right. \kern-\nulldelimiterspace} {\hslash c}\approx 1 
\mathord{\left/ {\vphantom {1 {137}}} \right. \kern-\nulldelimiterspace} 
{137}$ is the electromagnetic fine structure constant, $\alpha$ and $\alpha_{em} $ are the gravitational and electromagnetic coupling constants, 
$\alpha_{Q}$ and $\alpha_{a} $ are dimensionless constants characterizing the electromagnetic field source with the charge $Q$ and the ration of the  angular momentum $J$ to the mass $M$ in the Kerr and Kerr-Newman metrics.

For the Kerr-Newman metric, the quantities $\rho_{K}^{2}$ and $\Delta_{KN} $ in the 
dimensionless variables have the forms

\begin{equation}
\label{eq22}
\rho_{K}^{2} =\rho^{2}+\alpha_{a}^{2} \cos^{2}\theta ,
\end{equation}

\begin{equation}
\label{eq23}
\Delta_{KN} =\rho^{2}f_{KN} =\rho^{2}\left( {1-\frac{2\alpha }{\rho 
}+\frac{\alpha_{a}^{2} +\alpha_{Q}^{2} }{\rho^{2}}} \right).
\end{equation}
In the presence of outer and inner event horizons, $\alpha^{2}>\alpha 
_{a}^{2} +\alpha_{Q}^{2} $, and

\begin{equation}
\label{eq24}
\left( {\rho_{\pm } } \right)_{KN} =\alpha \pm \sqrt {\alpha^{2}-\alpha 
_{a}^{2} -\alpha_{Q}^{2} } .
\end{equation}
For an extremal Kerr-Newman field, $\alpha^{2}=\alpha_{a}^{2} +\alpha_{Q}^{2} 
,\,\,\,\left( {\rho_{+} } \right)_{KN}^{ext} =\left( {\rho_{-} } 
\right)_{KN}^{ext} =\alpha ,$ and

\begin{equation}
\label{eq25}
f_{RN}^{ext} =\frac{\left( {\rho -\alpha } \right)^{2}}{\rho^{2}}.
\end{equation}
For $\alpha^{2}<\alpha_{a}^{2} +\alpha_{Q}^{2} $, the case of a naked 
singularity of the Kerr-Newman field is realized.

For the Kerr field in formulas (\ref{eq23}) and (\ref{eq24}), $\alpha_{Q} =0$. For the 
Reissner-Nordstr\"{o}m field in (\ref{eq22}) - (\ref{eq24}), $\alpha_{a} =0$. 

\subsection{Motion of scalar particle.}
For uncharged  particles with zero-spin, the second-order equation in a curved space-time has the form

\begin{equation}
\label{eq26}
\left( {-g} \right)^{{-1} \mathord{\left/ {\vphantom {{-1} 2}} \right. 
\kern-\nulldelimiterspace} 2}\frac{\partial }{\partial x^{\mu }}\left[ 
{\left( {-g} \right)^{1 \mathord{\left/ {\vphantom {1 2}} \right. 
\kern-\nulldelimiterspace} 2}g^{\mu \nu }\frac{\partial }{\partial x^{\nu 
}}\Phi } \right]+m^{2}\Phi =0,
\end{equation}
where $g$ is the determinant of the metric. After separation of variables, the equation for the radial function $R\left( \rho \right)$ becomes

\begin{equation}
\label{eq27}
\frac{d^{2}R}{d\rho^{2}}+A\left( \rho \right)\frac{dR}{d\rho }+B\left( \rho 
\right)R=0,
\end{equation}
where $\rho =r \mathord{\left/ {\vphantom {r {l_{c} }}} \right. 
\kern-\nulldelimiterspace} {l_{c} }$.

We bring Eq. (\ref{eq27}) to the form of a  Schr\"{o}dinger equation with the 
effective potential $U_{eff} \left( \rho \right)$:

\begin{equation}
\label{eq28}
\bar{{R}}_{l} \left( \rho \right)=R\left( \rho \right)\exp \left[ \frac{1}{2}\int 
{A\left( {{\rho }'} \right)} d{\rho }'\right] ,
\end{equation}

\begin{equation}
\label{eq29}
\frac{d^{2}\bar{{R}}\left( \rho \right)}{d\rho^{2}}+2\left( {E_{Schr} 
-U_{eff} \left( \rho \right)} \right)\bar{{R}}\left( \rho \right)=0,
\end{equation}

\begin{equation}
\label{eq30}
U_{eff} \left( \rho \right)=E_{Schr} +\frac{1}{4}\frac{dA}{d\rho 
}+\frac{1}{8}A^{2}-\frac{1}{2}B,
\end{equation}

\begin{equation}
\label{eq31}
E_{Schr} =\frac{1}{2}\left( {\varepsilon^{2}-1} \right).
\end{equation}
The term $E_{Schr} $ given by  (\ref{eq31}) is distinguished in Eq. (\ref{eq29}) and at the 
same time added to (\ref{eq30}). This is done, on one hand, to give Eq.  (\ref{eq29}) the form of a Schr\"{o}dinger-type equation and, on the other hand, to ensure the classical asymptotic form of the effective potential as $\rho \to \infty $.

{\bf{3.2.1}} {\bf{Kerr and Kerr-Newman metrics}}. In this section, we use results in \cite{35}, 
where variables in Eq. (\ref{eq26}) were separates for the Kerr and Kerr-Newman metrics and for uncharged scalar particles.

In dimensionless variables (\ref{eq21}), the wave function has the form

\begin{equation}
\label{eq32}
\Phi_{KN} \left( {{\rm {\bm\rho }},t} \right)=R_{KN} \left( \rho 
\right)S\left( \theta \right)e^{-i\varepsilon t}e^{i\,m_{\varphi } \,\varphi 
},
\end{equation}
where $S\left( \theta \right)$ are oblate spheroidal harmonic functions $S_{lm_{\varphi } } \left( {ic,\cos \theta } \right)$, $c^{2}=\alpha_{a}^{2} \left( {\varepsilon^{2}-1} \right)$, and $l$ and $m_{\varphi 
} \,$ are the quantum numbers of the orbital momentum and its projection 
$\left( {\left| {m_{\varphi } } \right|\le l} \right)$.

The equations of the radial functions have the form \cite{35}

\begin{equation}
\label{eq33}
\begin{array}{c}
 \dfrac{d}{d\rho }\left( {\Delta_{KN} \dfrac{dR_{KN} }{d\rho }} 
\right)+\dfrac{1}{\Delta_{KN} }\left[ {\varepsilon^{2}\left( {\rho 
^{2}+\alpha_{a}^{2} } \right)^{2}-2\left( {2\alpha \rho -\alpha_{Q}^{2} } 
\right)\varepsilon \alpha_{a} m_{\varphi } +m_{\varphi }^{2} \alpha 
_{a}^{2} -} \right. \\ [10pt]
 \left. {-\left( {\varepsilon^{2}\alpha_{a}^{2} +\rho^{2}+\lambda 
_{lm_{\varphi } }^{KN} } \right)\Delta_{KN} } \right]R_{KN} =0, \\ 
 \end{array}
\end{equation}
where $\lambda_{lm_{\varphi } }^{KN} \left( {\alpha_{a} ,\varepsilon } 
\right)$ is the separation constant for Eq. (\ref{eq26}). In accordance with (\ref{eq27}) - (\ref{eq31}), for Eq. (\ref{eq33}), we can write

\begin{equation}
\label{eq34}
\begin{array}{c}
A_{KN} =\dfrac{2\left( {\rho -\alpha } \right)}{\Delta_{KN} }, \\ [10pt]
B_{KN} =\dfrac{1}{\Delta_{KN}^{2} }\left\{ {\left[ {\varepsilon \left( {\rho 
^{2}+\alpha_{a}^{2} } \right)-\alpha_{a} m_{\varphi } } \right]^{2}-\left( 
{\varepsilon^{2}\alpha_{a}^{2} -2\varepsilon \alpha_{a} m_{\varphi } } 
\right)\Delta_{KN} -\left( {\rho^{2}+\lambda_{lm_{\varphi } }^{KN} } 
\right)\Delta_{KN} } \right\}.
\end{array}
\end{equation}

In explicit form, the effective potential $U_{eff}^{KN} \left( \rho \right)$ in the Schr\"{o}dinger-type equation with (\ref{eq30}) and (\ref{eq34}) taken into account is

\begin{equation}
\label{eq36}
\begin{array}{c}
 U_{eff}^{KN} \left( \rho \right)=\dfrac{1}{2}\left( {\varepsilon^{2}-1} 
\right)+\dfrac{1}{2\Delta_{KN} }+\dfrac{\varepsilon^{2}\alpha_{a}^{2} 
-2\varepsilon \alpha_{a} m_{\varphi } }{2\Delta_{KN} }+\dfrac{\rho 
^{2}}{2\Delta_{KN} }+\dfrac{\lambda_{lm_{\varphi } }^{KN} }{2\Delta_{KN} 
}- \\ [10pt]
-\dfrac{\left( {\rho -\alpha } \right)^{2}}{2 \Delta_{KN}^{2} }
-\dfrac{1}{2\Delta_{KN}^{2} }\left[ {\varepsilon \left( {\rho^{2}+\alpha 
_{a}^{2} } \right)-\alpha_{a} m_{\varphi } } \right]^{2}. \\ 
 \end{array}
\end{equation}

{\bf{3.2.2}} {\bf{Asymptotic behavior of the effective potential}}. As $\rho \to \infty $, we have

\begin{equation}
\label{eq37}
\left. {U_{eff}^{KN} } \right|_{\rho \to \infty } =\,\frac{\alpha }{\rho 
}\left( {1-2\varepsilon^{2}} \right)+{\rm O}\left( {\frac{1}{\rho^{2}}} 
\right).
\end{equation}

As $\rho \to 0$,
\begin{equation}
\label{eq38}
\left. {U_{eff}^{KN} } \right|_{\rho \to 0} \to \mbox{const}_{KN} +{\rm 
O}\left( \rho \right).
\end{equation}

As $\rho \to \left( {\rho_{\pm } } \right)_{KN} $,
\begin{equation}
\label{eq39}
\left. {U_{eff}^{KN} } \right|_{\rho \to \left( {\rho_{\pm } } \right)_{KN} 
} =-\,\,\frac{1}{\left( {\rho -\left( {\rho_{\pm } } \right)_{KN} } 
\right)^{2}}\left\{ {\frac{1}{8}+\frac{\left[ {\varepsilon \left( {\left( 
{\rho_{\pm } } \right)_{KN}^{2} +\alpha_{a}^{2} } \right)-\alpha_{a} 
m_{\varphi } } \right]^{2}}{2\left( {\left( {\rho_{+} } \right)_{KN} 
-\left( {\rho_{-} } \right)_{KN} } \right)^{2}}} \right\}.
\end{equation}

For the Kerr metric, asymptotic formulas (\ref{eq37}) - (\ref{eq39}) hold with $\alpha_{Q} =0$. The asymptotic formulas for the Reissner-Nordstr\"{o}m metric can be obtained from (\ref{eq37}) and (\ref{eq39}) with $\alpha_{a} =0$. The asymptotic formulas for the Schwarzschild metric 
are obtained from (\ref{eq37}) and (\ref{eq39}) with $\alpha_{a} =0$ and $\alpha_{Q} =0$. Asymptotic formula (\ref{eq39}) indicates that the systems ''a scalar particle in Kerr, 
Kerr-Newman, Reissner-Nordstr\"{o}m, and  Schwarzschild fields'' are singular.

{\bf{3.2.3}} {\bf{Painlev\'{e}-Gullstrand metric in the coordinates $\left( {\bf T,r,\bm \theta ,\bm \varphi } \right)$}}. The stationary Eddington-Finkelstein \cite{36}, 
\cite{37} and Painlev\'{e}-Gullstrand 
\cite{38}, \cite{39} metrics and the  
nonstationary Lema\^{\i}tre -Finkelstein \cite{40}, 
\cite{37} and Kruskal-Szekeres 
\cite{41}, \cite{42} metrics were 
previously obtained by coordinate transformation of the Schwarzschild metric (\ref{eq8}) to eliminate the singularity on the event horizon $r_{0} $. 
But in quantum mechanics, the singularity for all mentioned metrics is manifested in the final results. In \cite{43}, this was shown in the example of a spin-1/2 
particle, and the equivalence of the Schwarzschild metric and metrics indicated above was shown for the regular solution $\varepsilon =0$. In this section, for the Painlev\'{e} -Gullstrand 
metric, we see the same behavior of the effective potential in a 
neighborhood of $\rho =2\alpha $ and $\rho \to \infty $ as for the Schwarzschild metric in Sec. 3.2.2 (see (\ref{eq37}), (\ref{eq39})) for $\alpha_{a} =0$ and $ \alpha_{Q} = 0$). 

In accordance with (\ref{eq11}), $\sqrt {-g} =r^{2}\sin \theta $,
\[
g^{00}=1,\,\,\,g^{01}=g^{10}=-\sqrt \frac{r_{0}} {r} ,\,\,\,g^{11}=-f_{S} 
,\,\,g^{22}=-\frac{1}{r^{2}},\,\,\,g^{33}=-\frac{1}{r^{2}\sin^{2}\theta }.
\]
We present the wave function $\Phi_{PG} \left( {{\rm {\bf r}},T} 
\right)$ for stationary states in the form

\begin{equation}
\label{eq40}
\Phi_{PG} \left( {{\rm {\bf r}},T} \right)=P_{l} \left( r 
\right)Y_{lm_{\varphi } } \left( {\theta ,\varphi } \right)e^{-iET}.
\end{equation}
After separation of variables in (\ref{eq26}) and substitution (\ref{eq40}), the 
equations for the radial functions $P_{l} \left( r \right)$ become

\begin{equation}
\label{eq41}
\begin{array}{c}
\dfrac{d^{2}P_{l} }{dr^{2}}+\left( {\dfrac{1}{r^{2}}\left( {2r-r_{0} } 
\right)-i2\sqrt {\dfrac{r_{0} }{r}} E} \right)\dfrac{1}{f_{S} }\dfrac{dP_{l} 
}{dr}+ \\ [10pt]
+\left[ {\dfrac{\left( {E^{2}-m^{2}} \right)}{f_{S} }-\dfrac{l\left( 
{l+1} \right)}{r^{2}f_{S} }-\dfrac{iE}{r^{2}f_{S} }\sqrt {\dfrac{r_{0} }{r}} 
\left( {2r-\dfrac{r_{0} }{2}} \right)} \right]P_{l} =0.
\end{array}
\end{equation}
By analogy with Eq. (\ref{eq27}), after passing to dimensionless variables (\ref{eq21}), we define

\begin{equation}
\label{eq42}
\begin{array}{l}
A_{PG} =\dfrac{1}{f_{S} }\left( {\dfrac{2\left( {\rho -\alpha } \right)}{\rho 
^{2}}-i2\varepsilon \sqrt {\dfrac{2\alpha }{\rho }} } \right), \\ [10pt]
B_{PG} =\frac{1}{f_{S} }\left( {\varepsilon^{2}-1} \right)-\dfrac{1}{\rho 
^{2}f_{S} }l\left( {l+1} \right)-i\varepsilon \dfrac{1}{\rho^{2}f_{S} }\sqrt 
{\dfrac{2\alpha }{\rho }} \left( {2\rho -\alpha } \right).
\end{array}
\end{equation}
Further, we bring Eq. (\ref{eq41}) to the form of a  Schr\"{o}dinger equation 
with the effective potential $U_{eff}^{PG} \left( \rho \right)$:

\begin{equation}
\label{eq44}
\begin{array}{l}
\bar{{P}}_{l} \left( \rho \right)=P_{l} \left( \rho \right)\exp 
\left[ \dfrac{1}{2}\int {A_{PG} \left( {{\rho }'} \right)} d{\rho }'\right] , \\ [10pt]
\dfrac{d^{2}\bar{{P}}_{l} \left( \rho \right)}{d\rho^{2}}+2\left( {E_{Schr} 
-U_{eff}^{PG} \left( \rho \right)} \right)\bar{{P}}_{l} \left( \rho 
\right)=0, \\[10pt]
U_{eff}^{PG} \left( \rho \right)=E_{Schr} +\dfrac{1}{4}\dfrac{dA_{PG} }{d\rho 
}+\dfrac{1}{8}A_{PG}^{2} -\dfrac{1}{2}B_{PG} , \\ [10pt]
E_{Schr} =\frac{1}{2}\left( {\varepsilon^{2}-1} \right).
\end{array}
\end{equation}
An explicit formula for $U_{eff}^{PG} \left( \rho \right)$ is given in Appendix A. Equalities (\ref{eq41}) - (\ref{eq44}) are complex. But analysis of the asymptotic behavior of $U_{eff}^{PG} \left( \rho \right)$ shows that the leading singularities are on the real axis and completely coincide with the 
singularities for the Schwarzschild metric. Therefore, the system ''a scalar particle in the Painlev\'{e} -Gullstrand field'' is singular. Obviously, we can also draw the same conclusion for the stationary Eddington-Finkelstein metric \cite{36}, \cite{37}.

{\bf{3.2.4}} {\bf{Extremal Kerr and Kerr-Newman fields}}. In the case of the extremal Kerr and Kerr-Newman fields, there exists a single event horizon $\rho =\alpha $. Moreover, $\alpha 
^{2}=\alpha_{a}^{2} +\alpha_{Q}^{2} $ and $\Delta_{KN} =\left( {\rho -\alpha 
} \right)^{2}$.

The expression for effective potential (\ref{eq36}) becomes

\begin{equation}
\label{eq48}
\begin{array}{c}
U_{eff}^{ext} =\dfrac{1}{2}\left( {\varepsilon^{2}-1} \right)+\dfrac{\rho 
^{2}}{2\left( {\rho -\alpha } \right)^{2}}+\frac{\lambda_{ext} }{2\left( 
{\rho -\alpha } \right)^{2}}+\dfrac{\varepsilon^{2}\alpha_{a}^{2} 
-2\varepsilon \alpha_{a} m_{\varphi } }{2\left( {\rho -\alpha } 
\right)^{2}}- \\ [10pt]
-\dfrac{\left( {\rho^{2}+\alpha_{a}^{2} } \right)^{2}}{2\left( 
{\rho -\alpha } \right)^{4}}\left( {\varepsilon -\dfrac{\alpha_{a} 
m_{\varphi } }{\rho^{2}+\alpha_{a}^{2} }} \right)^{2}.
\end{array}
\end{equation}
As $\rho \to \infty $, the asymptotic formula for $U_{eff}^{ext} $ remains as in (\ref{eq37}). As $\rho \to 0$, we must take the equality $\alpha ^{2}=\alpha_{a}^{2} +\alpha_{Q}^{2} $  into account in  $\mbox{const}_{KN} $ in asymptotic formula (\ref{eq38}) . As $\rho \to \alpha $ and under the condition that $\varepsilon \ne \varepsilon^{ext}=\alpha 
_{a} m_{\varphi } / (\alpha^{2}+\alpha_{a}^{2} )$, the effective potential 
has the form

\begin{equation}
\label{eq49}
\left. {U_{eff}^{ext} } \right|_{\rho \to \alpha } =-\,\frac{\left( {\alpha 
^{2}+\alpha_{a}^{2} } \right)^{2}}{2\left( {\rho -\alpha } 
\right)^{4}}\left( {\varepsilon -\varepsilon^{ext}} \right)^{2}+{\rm 
O}\left( {\frac{1}{\left( {\rho -\alpha } \right)^{2}}} \right).
\end{equation}
As $\rho \to \alpha $ and for $\varepsilon =\varepsilon^{ext}$, we can write the expression 
for $U_{eff}^{ext} $ as

\begin{equation}
\label{eq50}
\left. {U_{KN}^{ext} \left( {\varepsilon =\varepsilon^{ext}} \right)} 
\right|_{\rho \to \alpha } =\frac{1}{2\left( {\rho -\alpha } 
\right)^{2}}\left[ {C_{ext} +\lambda_{ext} +\alpha^{2}-\Omega^{2}} 
\right],
\end{equation}
where

\begin{equation}
\label{eq51}
C_{ext} =\varepsilon^{ext}\alpha_{a} \left( {\varepsilon^{ext}-2\alpha 
m_{\varphi } } \right),\,\,\,\,\,\Omega =\frac{2\alpha \alpha_{a} m_{\varphi } 
}{\left( {\alpha^{2}+\alpha_{a}^{2} } \right)}.
\end{equation}

If

\begin{equation}
\label{eq52}
C_{ext} +\lambda_{ext} +\alpha^{2}-\Omega^{2}>0
\end{equation}
in (\ref{eq50}), then there exists a potential barrier on the event horizon. If   \mbox{$C_{ext} 
	+\lambda_{ext} +\alpha^{2}-\Omega^{2}\ge 3/4$}, then the barrier becomes quantum 
mechanically impenetrable.

If 

\begin{equation}
\label{eq53}
-\dfrac{1}{4} <C_{ext} +\lambda_{ext} +\alpha^{2}-\Omega^{2}<0,
\end{equation}
then there exists a potential well near the event horizon, and in this case, the stationary bound state of scalar particles with $\varepsilon 
=\varepsilon^{ext}$ can exist in domains of the wave functions $\rho 
\in \left[ {\alpha ,\infty } \right)$ and $\rho \in \left( {0,\alpha } 
\right]$.

If

\begin{equation}
\label{eq54}
C_{ext} +\lambda_{ext} +\alpha^{2}-\Omega^{2}\le -\dfrac{1}{4},
\end{equation}
then the regime of particles ''falling'' on the event horizon is realized.

For $\varepsilon \ne \varepsilon^{ext}$, in accordance with asymptotic formula
(\ref{eq49}), the systems ''a scalar particle in an extremal Kerr and Kerr-Newman 
field'' are singular. For $\varepsilon =\varepsilon^{ext}$, the systems 
are singular if inequality (\ref{eq54}) is satisfied.

\subsection{Photon in gravitational and electromagnetic fields of asymptotically flat vacuum solutions  of General Relativity}

{\bf{3.3.1}} {\bf{Photon in a Kerr field}}. Teukolsky \cite{45} separated the variables in the Maxwell equations in a Kerr space-time \cite{44} (also see (\ref{eq15}), (\ref{eq18}), (\ref{eq19})). The function $\psi $ of the master equation was presented in the form

\begin{equation}
\label{eq55}
\psi =e^{-i\omega \,t}e^{im\,\,\varphi }S\left( \theta \right)R\left( r 
\right).
\end{equation}
The function $\psi$ in  \cite{45} is related to the components of the electromagnetic field tensor convoluted with components of the Kinnersley 
tetrad \cite{46} in the Newman-Penrose formalism 
\cite{47}. For our analysis, the Teukolsky separation of variables has a significant disadvantage: the equation for the radial function $R\left( r \right)$ and hence also the effective potential in the 
Schr\"{o}dinger equation are complex.

Lunin \cite{48} introduced new more complicated relation of ansatz (\ref{eq55}) to the components of the electromagnetic field potential $A^{\mu }\left( {{\rm {\bf r}},t} \right)$. As a result, a separation of variables with a real equation for the 
radial function $R\left( r \right)$ was produced. In the Lunin terminology, variables are separated for the ''electric polarization'' and for the ''the magnetic polarization'' (Teukolsky did not separate the variables for ''the magnetic polarization''). For a better understanding, we write the final formulas from \cite{48}.

The electrical polarization is described by the relations
\begin{equation}
\label{eq56}
\begin{array}{c}
 l_{\pm }^{\mu } A_{\mu }^{\left( {el} \right)} =\pm \dfrac{r}{1\pm i\mu 
r}\hat{{l}}_{\pm } \Psi ,\,\,\,\,\,m_{\pm }^{\mu } A_{\mu }^{\left( {el} 
\right)} =\mp \dfrac{iac_{\theta } }{1\pm i\mu ac_{\theta } }\hat{{m}}_{\pm 
} \Psi ,\,\,\,\,\Psi =e^{-i\,\omega \,t+i\,m\,\,\varphi }R\left( r 
\right)S\left( \theta \right), \\ [10pt]
 \dfrac{E_{\theta } }{s_{\theta } }\dfrac{d}{d\theta }\left[ {\dfrac{s_{\theta 
} }{E_{\theta } }\dfrac{d}{d\theta }S} \right]+\left\{ {-\dfrac{2\Lambda 
}{E_{\theta } }+\left( {a\omega c_{\theta } } 
\right)^{2}-\dfrac{m^{2}}{s_{\theta }^{2} }-C} \right\}S=0, \\ [10pt]
 E_{r} \dfrac{d}{dr}\left[ {\dfrac{\Delta_{K} }{E_{r} }\dfrac{d}{dr}R} 
\right]+\left\{ {\dfrac{2\Lambda }{E_{r} }+\left( {\omega r} 
\right)^{2}+\dfrac{\left( {am} \right)^{2}}{\Delta_{K} }+\dfrac{r_{0} r\omega 
^{2}\Delta_{0} }{\Delta_{K} }-\dfrac{2r_{0} ar\omega m}{\Delta_{K} }+C} 
\right\}R=0, \\ 
 \end{array}
\end{equation}
where
$ E_{r} =1+\left( {\mu r} \right)^{2},\,\,\,E_{\theta } =1-\left( {\mu 
ac_{\theta } } \right)^{2},\,\,\,\Delta_{K} =r^{2}+a^{2}-r_{0} 
r,\,\,\,\Delta_{0} =r^{2}+a^{2},\,\,s_{\theta } ,\,\,c_{\theta }=\sin 
\theta ,\newline \cos \theta , \,\,\, \xi =a\mu \left[ {m-a\omega +\omega / a\mu^{2}} 
\right],\,\,\,C=-\xi -2am\omega +\left( {a\omega } \right)^{2}$, $\mu$ is the separation constant,

\begin{equation}
\label{eq57}
\begin{array}{c}
 l_{+}^{\mu } =l^{\mu } ,\,\,\,\,\,l_{-}^{\mu } =-2\rho_{k}^{2} n^{\mu } /
\Delta_{K},\\ [10pt]
m_{+}^{\mu } =\sqrt 2 \rho m^{\mu } 
,\,\,\,\,m_{-}^{\mu } =\sqrt 2 \rho^{\ast }m^{\ast \mu },\,\,\,\,\, \rho =r+ia\cos \theta ,\\ [10pt]
 l_{\pm }^{\mu } \partial_{\mu } =\partial_{r} \pm \left[ 
{\dfrac{r^{2}+a^{2}}{\Delta_{K} }\partial_{t} +\dfrac{a}{\Delta_{K} 
}\partial_{\varphi } } \right],\,\,\,\,\,\,m_{\pm }^{\mu } \partial_{\mu } 
=\partial_{\theta } \pm \left[ {ia\sin \theta \,\partial_{t} 
+\dfrac{i}{\sin \theta }\partial_{\varphi } } \right], \\ [10pt]
 \end{array}
\end{equation}
$l^{\mu } ,\,\,n^{\mu },\,\,m^{\mu }$, and $m^{\ast \mu}$ are components of the Kinnersley tetrad vectors \cite{46} with $l_{\mu } n^{\mu }=-1$ and $m_{\mu } m^{\ast \mu }=1$.

The magnetic polarization is described by the formulas

\begin{equation}
\label{eq58}
\begin{array}{c}
 l_{\pm }^{\mu } A_{\mu }^{\left( {mgn} \right)} =\pm \dfrac{ia}{1\pm i\mu 
a}\hat{{l}}_{\pm } \Psi ,\,\,\,\,\,m_{\pm }^{\mu } A_{\mu }^{\left( {mgn} 
\right)} =\mp \dfrac{1}{c_{\theta } \mp \mu }\hat{{m}}_{\pm } \Psi 
,\,\,\,\,\,\Psi =e^{-i\,\omega \,t+i\,m\,\,\varphi }R\left( r \right)S\left( 
\theta \right), \\ [10pt]
 \dfrac{M_{\theta } }{s_{\theta } }\dfrac{d}{d\theta }\left[ {\dfrac{s_{\theta 
} }{M_{\theta } }\dfrac{d}{d\theta }S} \right]+\left\{ 
{-\dfrac{m^{2}}{s_{\theta }^{2} }-\dfrac{2\Lambda }{M_{\theta } }+\left( 
{a\omega c_{\theta } } \right)^{2}-C} \right\}S=0, \\ [10pt]
 M_{r} \dfrac{d}{dr}\left[ {\dfrac{\Delta_{K} }{M_{r} }\dfrac{d}{dr}R} 
\right]+\left\{ {-\dfrac{2\Lambda a^{2}}{M_{r} }+\dfrac{\left( {am} 
\right)^{2}}{\Delta_{K} }+\left( {r\omega } \right)^{2}+\dfrac{r_{0} r\omega 
^{2}\Delta_{0} }{\Delta_{K} }-\dfrac{2r_{0} ar\omega m}{\Delta_{K} }+C} 
\right\}R=0, \\ 
 \end{array}
\end{equation}
where $M_{r} =r^{2}+\mu^{2}a^{2},\,\,M_{\theta } =c_{\theta }^{2} -\mu 
^{2},\,\,\Lambda =\mu \left[ {-a\omega +m+a\omega \mu^{2}} 
\right]$, and $C=\Lambda \mathord{\left/ {\vphantom {\Lambda {\mu^{2}}}} 
\right. \kern-\nulldelimiterspace} {\mu^{2}}-a\omega \left[ {-a\omega +2m} 
\right]$.

{\bf{Asymptotic behavior of the effective potential $\bf U_{eff}^{K} \left( r \right)$}}. We can obtain Schr\"{o}dinger-type equations for the functions $\bar{{R}}^{\left( e \right)}\left( r \right)$ and $\bar{{R}}^{\left( m \right)}\left( r \right)$  with the effective potentials $U_{eff}^{\left( e 
	\right)} \left( r \right)$ and $U_{eff}^{\left( m \right)} \left( r 
\right)$ from the equations for the radial function $R\left( r \right)$ in (\ref{eq56}) and 
(\ref{eq58}).

The asymptotic formulas for the potentials $U_{eff}^{\left( e \right)} \left( r 
\right)$ and $U_{eff}^{\left( m \right)} \left( r \right)$ are the same as $r\to \infty ,\,\,r\to 0$, and $r\to \left( {r_{\pm } } \right)_{K} $. We further analyze $U_{eff}^{\left( e \right)} \left( r 
\right)$:

\begin{equation}
\label{eq59}
\begin{array}{c}
 U_{eff}^{\left( e \right)} =\dfrac{1}{2}\omega^{2}+\dfrac{1}{2\Delta_{K} 
}-\dfrac{2\mu^{2}}{E_{r}^{2} }\left( {1-\mu^{2}r^{2}} \right)-\dfrac{\mu 
^{2}r\left( {2r-r_{0} } \right)}{2E_{r} \Delta_{K} 
}+\dfrac{1}{2}\dfrac{\mu^{4}r^{2}}{E_{r}^{2} }-\dfrac{\Lambda }{E_{r} 
\Delta_{K} }-\dfrac{C}{2}- \\ [10pt]
 -\dfrac{1}{8}\dfrac{\left( {2r-r_{0} } \right)^{2}}{\Delta_{K}^{2} 
}-\cfrac{\left[ {\omega \left( {r^{2}+a^{2}} \right)-am} \right]^{2}-\left( 
{\omega^{2}a^{2}-2\omega am} \right)\Delta_{K} }{2\Delta_{K}^{2} }. \\ 
 \end{array}
\end{equation}
We have 

\begin{equation}
\label{eq60}
\left. {U_{eff}^{\left( e \right)} } \right|_{r\to \infty } 
=\,-\,\,\frac{\omega^{2} r_{0}}{r}
\end{equation}
as $r\to \infty $,

\begin{equation}
\label{eq61}
\left. {U_{eff}^{\left( e \right)} } \right|_{r\to 0} =\mbox{const}+{\rm 
O}\left( \rho \right)
\end{equation}
as $r\to 0$, and 

\begin{equation}
\label{eq62}
\left. {U_{eff}^{\left( e \right)} } \right|_{r\to \left( {r_{\pm } } 
\right)_{K} } =-\frac{1}{\left( {r-\left( {r_{\pm } } \right)_{K} } 
\right)^{2}}\left[ {\frac{1}{8}+\frac{\left( {\omega^{2}\left( {\left( 
{r_{\pm } } \right)_{K}^{2} +a^{2}} \right)-am} \right)^{2}}{2\left( {\left( 
{r_{+} } \right)_{K} -\left( {r_{-} } \right)_{K} } \right)^{2} }} 
\right]+{\rm O}\left( {\frac{1}{\left| {r-\left( {r_{\pm } } \right)_{K} } 
\right|}} \right).
\end{equation}
as $r\to \left( {r_{\pm } } \right)_{k} $.

For the asymptotic of effective potentials $U_{eff}^{\left( e \right)} 
\left( r \right)$ and $U_{eff}^{\left( m \right)} \left( r \right)$ in 
Schr\"{o}dinger-type equations, the regime of particle 
''falling'' on the event horizons is realized (see (\ref{eq62})):

\begin{equation}
\label{eq63}
\left. {\bar{{R}}^{\left( e \right)}} \right|_{r\to \left( {r_{\pm } } 
\right)_{K} } ,\,\,\,\left. {\bar{{R}}^{\left( m \right)}} \right|_{r\to 
\left( {r_{\pm } } \right)_{K} } \sim \left( {r-\left( {r_{\pm } } 
\right)_{K} } \right)^{1 \mathord{\left/ {\vphantom {1 2}} \right. 
\kern-\nulldelimiterspace} 2}\sin \theta \left( {\sqrt K \ln \left( 
{r-\left( {r_{\pm } } \right)_{K} } \right)+\varphi_{0} } \right),
\end{equation}
where $\varphi_{0} $ is an arbitrary phase.

Functions (\ref{eq63}) have an unbounded number of zeros as $r\to \left( 
{r_{\pm } } \right)_{K} $. It can be seen from the relations in (\ref{eq56}) and (\ref{eq58}) that the electromagnetic potentials $A_{\mu }^{\left( {el} 
\right)} \left( {{\rm {\bf r}},t} \right)$ and $A_{\mu }^{\left( {mgn} 
\right)} \left( {{\rm {\bf r}},t} \right)$ also oscillate in neighborhoods of event 
horizons as do the radial functions $\bar{{R}}^{\left( 
e \right)}\left( r \right)$ and $\bar{{R}}^{\left( m \right)}\left( r \right)$ 
in (\ref{eq63}). We can conclude that  the system ''a photon in the Kerr gravitational field'' is singular after quantization of the electromagnetic field.

{\bf{3.3.2}} {\bf{Photon in a Kerr-Newman field}}. Variables in the Maxwell equations in a Kerr-Newman space-time can be separated using Lunin's work \cite{48} for the Kerr geometry. For fermions, Chandrasekhar's paper \cite{49} was similarly generalized previously for the Kerr geometry by Page \cite{50}.

We must first change 

\begin{equation}
\label{eq64}
\Delta_{K} \to \Delta_{KN} =r^{2}+a^{2}-\left( {r_{0} r-r_{Q}^{2} } 
\right)
\end{equation}
in the components of the Kinnersley tetrad in (\ref{eq57}). Following the Lunin formalism, we can then obtain formulas (\ref{eq56}) and (\ref{eq58}) 
and the effective potential $U_{KN}^{\left( e \right)} \left( r \right)$ 
given by (\ref{eq59}) with the changes $\Delta_{K} \to \Delta_{KN} $ and $r_{0} r\to 
\left( {r_{0} r-r_{Q}^{2} } \right)$ in them. In this case, the angular 
equations in (\ref{eq56}) and (\ref{eq58}) remain unchanged.

The asymptotic formulas for $U_{KN}^{\left( e \right)} \left( r 
\right)$ and $U_{KN}^{\left( m \right)} \left( r \right)$ are the same for $r\to 
\infty ,\,\,r\to 0$, and $r\to \left( {r_{\pm } } \right)_{KN} $. We have

\begin{equation}
\label{eq65}
\left. {U_{KN}^{\left( e \right)} } \right|_{r\to \infty } 
= -\frac{\omega^{2} r_{0} }{r}
\end{equation}
as $r\to \infty$, 

\begin{equation}
\label{eq66}
\left. {U_{KN}^{\left( e \right)} } \right|_{r\to 0} =\left( {\mbox{const}} 
\right)_{KN} +{\rm O}\left( \rho \right)
\end{equation}
as $r\to 0$, and

\begin{equation}
\label{eq67}
\begin{array}{c}
\left. {U_{KN}^{\left( e \right)} } \right|_{r\to \left( {r_{\pm } } 
\right)_{KN} } =-\dfrac{1}{\left( {r-\left( {r_{\pm } } \right)_{KN} } 
\right)^{2}}\left[ {\frac{1}{8}+\dfrac{\left( {\omega \left( {\left( {r_{\pm 
} } \right)_{KN}^{2} +a^{2}} \right)-am} \right)^{2}}{2\left( {\left( {r_{+} 
} \right)_{KN} -\left( {r_{-} } \right)_{KN} } \right)^{2} }} 
\right]+\\ [10pt]
+{\rm O}\left( {\dfrac{1}{\left| {r-\left( {r_{\pm } } \right)_{KN} } 
\right|}} \right)
\end{array}
\end{equation}
as $r\to \left( {r_{\pm } } \right)_{KN} $. In (\ref{eq67}), the radii of the outer and inner event horizons of the Kerr-Newman metric are defined in (\ref{eq17}). In accordance with (\ref{eq67}), repeating the arguments in Sec. 3.3.1, we can conclude that the system ''photon in a Kerr-Newman field'' is singular after quantization of the electromagnetic field. The singularity of the system ''a photon in Reissner-Nordstr\"{o}m and Schwarzschild fields'' also follows from (\ref{eq62}) and (\ref{eq67}).

{\bf{3.3.3}} {\bf{Extremal Kerr-Newman field}}. In this case, there exists a unique event horizon with $r={r_{0}/2 }$. Moreover, $a^{2}+r_{Q}^{2} ={r_{0}^{2}/4 }$ and $\Delta_{KN}^{ext} =\left( r - \left(r_0 /2 \right)  \right) ^2$.

The effective potential $U_{KN}^{ext} \left( r \right)$ in a
Schr\"{o}dinger-type equation for the radial function $\bar{{R}}_{KN}^{\left( e \right)} \left( r \right)$ has the form

\begin{equation}
\label{eq68}
\begin{array}{c}
 U_{KN}^{ext} =\dfrac{1}{2}\omega^{2}-\dfrac{2\mu^{2}\left( {1-\mu 
^{2}r^{2}} \right)}{E_{r}^{2} }-\dfrac{\mu^{2}r}{E_{r}^{2} \left( {r-\left( 
{{r_{0} } \mathord{\left/ {\vphantom {{r_{0} } 2}} \right. 
\kern-\nulldelimiterspace} 2} \right)} \right)}+\dfrac{1}{2}\dfrac{\mu 
^{4}r^{2}}{E_{r}^{2} }-\dfrac{\xi }{E_{r} \left( {r-\left( {{r_{0} 
} \mathord{\left/ {\vphantom {{r_{0} } 2}} \right. 
\kern-\nulldelimiterspace} 2} \right)} \right)^{2}}- \\ [10pt]
 -\dfrac{\left[ {\omega \left( {r^{2}+a^{2}} \right)-am} \right]^{2}}{2\left( 
{r-\left( {{r_{0} } \mathord{\left/ {\vphantom {{r_{0} } 2}} \right. 
\kern-\nulldelimiterspace} 2} \right)} \right)^{4}}+\dfrac{\omega 
^{2}a^{2}-\omega am}{2\left( {r-\left( {{r_{0} } \mathord{\left/ 
{\vphantom {{r_{0} } 2}} \right. \kern-\nulldelimiterspace} 2} \right)} 
\right)^{2}}. \\ 
 \end{array}
\end{equation}
The asymptotic formulas for $U_{KN}^{ext} $ as $r\to \infty $ and as $r\to 0$ remain as in (\ref{eq65}) and (\ref{eq66}) with the equality $a^{2}+r_{Q}^{2} ={r_{0}^{2} } \mathord{\left/ {\vphantom {{r_{0}^{2} } 4}} \right. \kern-\nulldelimiterspace} 4$ taken into account in  (\ref{eq66}).

In a neighborhood of the unique event horizon,

\begin{equation}
\label{eq69}
\left. {U_{KN}^{ext} } \right|_{r\to {r_{0} } \mathord{\left/ {\vphantom 
{{r_{0} } 2}} \right. \kern-\nulldelimiterspace} 2} =-\frac{\left[ {\omega 
\left( {\left( {{r_{0}^{2} } \mathord{\left/ {\vphantom {{r_{0}^{2} } 4}} 
\right. \kern-\nulldelimiterspace} 4} \right)+a^{2}} \right)-am} 
\right]^{2}}{2\left( {r-\left( {{r_{0} } \mathord{\left/ {\vphantom 
{{r_{0} } 2}} \right. \kern-\nulldelimiterspace} 2} \right)} 
\right)^{4}}.
\end{equation}
For $\omega \ne {am} \mathord{\left/ {\vphantom {{am} {\left( {\left( 
{{r_{0}^{2} } \mathord{\left/ {\vphantom {{r_{0}^{2} } 4}} \right. 
\kern-\nulldelimiterspace} 4} \right)+a^{2}} \right)}}} \right. 
\kern-\nulldelimiterspace} {\left( {\left( {{r_{0}^{2} } \mathord{\left/ 
{\vphantom {{r_{0}^{2} } 4}} \right. \kern-\nulldelimiterspace} 4} 
\right)+a^{2}} \right)}$, the system ''a photon in an extremal 
Kerr-Newman field'' is singular. 
For $\omega^{ext}={am} \mathord{\left/ {\vphantom {{am} {\left( {\left( 
{{r_{0}^{2} } \mathord{\left/ {\vphantom {{r_{0}^{2} } 4}} \right. 
\kern-\nulldelimiterspace} 4} \right)+a^{2}} \right)}}} \right. 
\kern-\nulldelimiterspace} {\left( {\left( {{r_{0}^{2} } \mathord{\left/ 
{\vphantom {{r_{0}^{2} } 4}} \right. \kern-\nulldelimiterspace} 4} 
\right)+a^{2}} \right)}$, the leading singularity of the potential in a neighborhood of the event horizon with $r={r_{0} } \mathord{\left/ 
{\vphantom {{r_{0} } 2}} \right. \kern-\nulldelimiterspace} 2$ has the 
form

\begin{equation}
\label{eq70}
\begin{array}{l}
 \left. {U_{eff}^{ext} } \right|_{r\to {r_{0} } \mathord{\left/ {\vphantom 
{{r_{0} } 2}} \right. \kern-\nulldelimiterspace} 2} =\dfrac{1}{2\left( 
{r-\left( {{r_{0} } \mathord{\left/ {\vphantom {{r_{0} } 2}} \right. 
\kern-\nulldelimiterspace} 2} \right)} \right)^{2}}\left[ 
{-\dfrac{a^{2}m^{2}r_{0}^{2} }{\left( {\left( {{r_{0}^{2} } \mathord{\left/ 
{\vphantom {{r_{0}^{2} } 4}} \right. \kern-\nulldelimiterspace} 4} 
\right)+a^{2}} \right)^{2}}+\dfrac{a^{4}m^{2}}{\left( {\left( {{r_{0}^{2} } 
\mathord{\left/ {\vphantom {{r_{0}^{2} } 4}} \right. 
\kern-\nulldelimiterspace} 4} \right)+a^{2}} \right)^{2}}-} \right. \\ [10pt]
 \left. {-\dfrac{a^{2}m^{2}}{\left( {{r_{0}^{2} } \mathord{\left/ {\vphantom 
{{r_{0}^{2} } 4}} \right. \kern-\nulldelimiterspace} 4} 
\right)+a^{2}}-\dfrac{2\xi }{E_{r} }} \right]=\dfrac{N_{ext} }{2\left( 
{r-\left( {{r_{0} } \mathord{\left/ {\vphantom {{r_{0} } 2}} \right. 
\kern-\nulldelimiterspace} 2} \right)} \right)^{2}}. \\ 
 \end{array}
 \end{equation}
 If $N_{ext} >0$, then there exists a potential barrier on the event horizon. If 
$N_{ext} \ge 3 \mathord{\left/ {\vphantom {3 4}} \right. 
\kern-\nulldelimiterspace} 4$, then this barrier is impermeable. If $-\,1 
\mathord{\left/ {\vphantom {1 4}} \right. \kern-\nulldelimiterspace} 
4<N_{ext} <0$, then there exists a potential well with possible realization of stationary photon states. If $N_{ext} \le -\,1 \mathord{\left/ {\vphantom {1 
4}} \right. \kern-\nulldelimiterspace} 4$, then the regime ofa  photon "falling" on the event 
horizon is realized.

Extremal Kerr and Reissner-Nordstr\"{o}m fields can be easily analyzed analogously.

\subsection{Spin-1/2 particle in Schwarzschild, Reissner-Nordstr\"{o}m, Kerr, and Kerr-Newman gravitational fields}

Stationary states of spin-1/2 particles were studied in \cite{4} - \cite{6} using second-order equations with effective potentials. Below, we briefly present the results of thise research.

{\bf{3.4.1}} {\bf{Asymptotic behavior of the effective potential}}. 
As $\rho \to \infty $ for the most general Kerr-Newman metric,

\begin{equation}
\label{eq71}
\left. {U_{eff}^{KN} } \right|_{\rho \to \infty } =\left( {1-2\varepsilon 
^{2}} \right)\frac{\alpha }{\rho }+\frac{\alpha_{em} \varepsilon }{\rho 
}+{\rm O}\left( {\frac{1}{\rho^{2}}} \right).
\end{equation}
For the Kerr and Schwarzschild metrics in (\ref{eq71}), $\alpha_{em} =0$.

As $\rho \to 0$, 

\begin{equation}
\label{eq72}
\left. {U_{eff}^{S} } \right|_{\rho \to 0} =\frac{5}{32\rho^{2}}
\end{equation}
for the Schwarzschild metric,

\begin{equation}
\label{eq73}
\left. {U_{eff}^{RN} } \right|_{\rho \to 0} =\frac{3}{8\rho^{2}}
\end{equation}
for the Reissner-Nordstr\"{o}m metric, and

\begin{equation}
\label{eq74}
\begin{array}{l}
\left. {U_{eff}^{K} } \right|_{\rho \to 0} ={\text{const}},\,\,\,
\left. {U_{eff}^{KN} } \right|_{\rho \to 0} ={\text{const}}
\end{array}
\end{equation}
for the Kerr and Kerr-Newman metric.

In the presence of an event horizon, for the Kerr-Newman metric and 
$\varepsilon \ne \varepsilon_{KN}^{st} $,

\begin{equation}
\label{eq75}
\left. {U_{eff}^{KN} \left( {\varepsilon \ne \varepsilon_{KN}^{st} } 
\right)} \right|_{\rho \to \left( {\rho_{\pm } } \right)_{KN} } 
=-\frac{1}{\left( {\rho -\left( {\rho_{\pm } } \right)_{KN} } 
\right)^{2}}\left[ {\frac{1}{8}+\frac{\left( {\varepsilon -\varepsilon 
_{KN}^{st} } \right)^{2}\left( {\left( {\rho_{\pm } } \right)_{KN}^{2} 
+\alpha_{a}^{2} } \right)^{2}}{2\left[ {\left( {\rho_{+} } \right)_{KN} 
-\left( {\rho_{-} } \right)_{KN} } \right]^{2}}} \right],
\end{equation}
where

\begin{equation}
\label{eq76}
\varepsilon_{KN}^{st} =\frac{\alpha_{a} m_{\varphi } +\alpha_{em} 
\left( {\rho_{\pm } } \right)_{KN} }{\alpha_{a}^{2} +\left( {\rho_{\pm } 
} \right)_{KN}^{2} }.
\end{equation}
 From (\ref{eq75}), (\ref{eq76}) and (\ref{eq22}) - (\ref{eq24}), we obtain the asymptotic formula in a neighborhood of the event horizon for the Kerr metric with $\alpha_{Q} 
=0$, for the Reissner-Nordstr\"{o}m metric with $\alpha_{a} 
=0$, and for the Schwarzschild metric with $\alpha_{Q} =0$ and $\alpha_{a} =0$.

For the extremal Kerr-Newman field with a unique event horizon \mbox{$\left( 
{\rho_{+} } \right)_{KN} =\left( {\rho_{-} } \right)_{KN} =\alpha $},

\begin{equation}
\label{eq77}
\left. {U_{eff}^{ext} \left( {\varepsilon \ne \varepsilon_{KN}^{ext} } 
\right)} \right|_{\rho \to \alpha } =-\frac{\left( {\alpha_{a}^{2} +\alpha 
^{2} } \right)^{2}\left( {\varepsilon -\varepsilon_{KN}^{ext} } 
\right)^{2}}{2\left( {\rho -\alpha } \right)^{4}},
\end{equation}
where 

\begin{equation}
\label{eq78}
\varepsilon_{KN}^{ext} =\frac{\alpha m_{\varphi } +\alpha_{em} \alpha 
}{\alpha_{a}^{2} +\alpha^{2} }.
\end{equation}
We can obtain analogous asymptotic formulas for extremal Kerr and 
Reissner-Nordstr\"{o}m fields from (\ref{eq77}) and (\ref{eq78}).

It was proved in \cite{4} - \cite{6} that there exist stationary bound states of spin-1/2 particles in the considered gravitational fields.

For the Kerr-Newman metrics in the presence of event horizons $\left( {\rho 
_{\pm } } \right)_{KN} $, the stationary state energies are given by expression (\ref{eq76}). We have 

\begin{equation}
\label{eq79}
\varepsilon_{K}^{st} =\frac{\alpha_{a} m_{\varphi } }{2\alpha \left( 
{\rho_{\pm } } \right)_{KN} }
\end{equation}
for the Kerr metric,

\begin{equation}
\label{eq80}
\varepsilon_{RN}^{st} ={\alpha_{em} } \mathord{\left/ {\vphantom 
{{\alpha_{em} } {\left( {\rho_{\pm } } \right)_{KN} }}} \right. 
\kern-\nulldelimiterspace} {\left( {\rho_{\pm } } \right)_{KN} }
\end{equation}
for the Reissner-Nordstr\"{o}m metric, and

\begin{equation}
\label{eq81}
\varepsilon_{S}^{st} =0
\end{equation}
for the Schwarzschild metric. 
The energy of a stationary bound state for the extremal Kerr-Newman field 
is given by expression (\ref{eq78}). We have          

\begin{equation}
\label{eq82}
\varepsilon_{K}^{ext} ={m_{\varphi } } \mathord{\left/ {\vphantom 
{{m_{\varphi } } {2\alpha }}} \right. \kern-\nulldelimiterspace} {2\alpha }
\end{equation}
for the Kerr metric, and

\begin{equation}
\label{eq83}
\varepsilon_{RN}^{ext} ={\alpha_{em} } \mathord{\left/ {\vphantom {{\alpha 
_{em} } \alpha }} \right. \kern-\nulldelimiterspace} \alpha 
\end{equation}
for the Reissner-Nordstr\"{o}m metric.
For all considered metrics, the asymptotic effective potential 
in a neighborhood of event horizons with energies of stationary states (\ref{eq76}) and  (\ref{eq79}) - (\ref{eq81}) have the same form

\begin{equation}
\label{eq84}
\left. {U_{eff} \left( {\varepsilon =\varepsilon^{st} } \right)} 
\right|_{\rho \to \rho_{\pm } } =-\frac{3}{32}\frac{1}{\left( {\rho -\rho 
_{\pm } } \right)^{2}}.
\end{equation}

Asymptotic formula (\ref{eq84}) admits the existence of stationary bound states of spin-1/2 particles. Expression (\ref{eq75}) does not coincide with asymptotic formula (\ref{eq84}) as 
\mbox{$\varepsilon \to 0$}. For their coincidence, in the expressions for $U_{eff} $ (see Appendix B), terms that are insignificant at a finite $\varepsilon$ but noticeably contribute to the coefficient with the leading singularity as $\varepsilon \to 0$ must be taken into account.

For the metrics of the Kerr, Kerr-Newman, and Reissner-Nordstr\"{o}m extremal fields, the 
asymptotic effective potential as $\rho \to \alpha $ with energies of stationary bound states  (\ref{eq78}), (\ref{eq82}) and (\ref{eq83}) has the form

\begin{equation}
\label{eq85}
\left. {U_{eff}^{ext} \left( {\varepsilon =\varepsilon_{KN}^{ext} } 
\right)} \right|_{\rho \to \alpha } =-\frac{1}{2\left( {\rho -\alpha } 
\right)^{2}}\left[ {\frac{1}{4}-\left( {\lambda^{2}+\alpha^{2}-\alpha 
^{4}\Omega^{2}} \right)} \right]+{\rm O}\left( {\frac{1}{\left| {\rho 
-\alpha } \right|}} \right),
\end{equation}
where

\[
\Omega =-\frac{m_{\varphi } \alpha_{a} +\alpha_{em} \alpha 
}{\alpha^{2} +\alpha_{a}^{2} }\frac{2\alpha_{a}^{2} }{\alpha^{3} 
}+\frac{2m_{\varphi } \alpha_{a} }{\alpha^{3} }+\frac{\alpha 
_{em} }{\alpha^{2} }
\]
and $\lambda \left( {\varepsilon ,\alpha_{a} 
,j,m_{\varphi } } \right)$ is the separation constant  in the Chandrasekhar--Page equations \cite{49}, \cite{50}.

We can write the condition for the existence of a potential well in potential (\ref{eq85}) and the condition 
for the existence of the stationary bound states with energies (\ref{eq78}), (\ref{eq82}), and (\ref{eq83}) in it as

\begin{equation}
\label{eq86}
0<\lambda^{2}+\alpha^{2}-\alpha^{4}\Omega^{2}<\dfrac{1}{4} \,\,\,(\text{Kerr-Newman metric}),\end{equation}

\begin{equation}
\label{eq87}
0<\lambda^{2}+\alpha^{2}-m_{\varphi }^{2} <\dfrac{1}{4},\,\,\,\,\Omega ={m_{\varphi } } 
\mathord{\left/ {\vphantom {{m_{\varphi } } {\alpha_{a}^{2} }}} \right. 
\kern-\nulldelimiterspace} {\alpha_{a}^{2} }\,\,\,(\text{Kerr metric}),
\end{equation}

\begin{equation}
\label{eq88}
0<\kappa^{2}+\alpha^{2}-\alpha_{em}^{2} <\dfrac{1}{4},\,\,\,\,\lambda =\kappa ,\,\,\Omega 
={\alpha_{em} } / {\alpha^{2}} \,\,\,\,\,(\text{Reissner-Nordstr\"{o}m metric}).
\end{equation}

Therefore, for all considered metrics, if $\varepsilon \ne \varepsilon 
^{st} $, then systems ''a spin 1/2  particle in gravitational fields with event horizons'' are singular. This statement also holds for extremal fields if $\varepsilon \ne \varepsilon^{ext} $.

There also exist regular stationary solutions $\varepsilon =\varepsilon 
^{st} $ given by (\ref{eq76}) and  (\ref{eq79}) - (\ref{eq81}) for metrics with event horizons and 
$\varepsilon =\varepsilon^{ext} $ given by  (\ref{eq78}) and (\ref{eq82}), (\ref{eq83}) under conditions 
(\ref{eq86}) - (\ref{eq88}) for extremal fields with a unique event horizons. 
Solutions $\varepsilon =\varepsilon^{st} $ correspond to square-integrable wave functions vanishing on event horizons. Particles in stationary bound states are located near event horizons (over outer and under inner event horizons) with a high probability. The probability density maximums for detecting particles are separated from event horizons by fractions of the Compton wavelength of bound fermions.

\subsection{Discussion of results in this section}

We present the final results for the leading singularities of 
effective potentials $U_{eff} \left( r \right)$ in neighborhoods of 
event horizons. The results for photons are given in natural units.

Schwarzschild field:

\begin{itemize}
\item scalar particle and fermion with $\varepsilon \ne \varepsilon_{S}^{st} $ (see (\ref{eq39})),
\end{itemize}

\[
\left. {U_{eff}^{s} } \right|_{\rho \to 2\alpha } =-\dfrac{1}{\left( {\rho 
-2\alpha } \right)^{2}}\left( {\dfrac{1}{8}+2\alpha^{2}\varepsilon^{2}} 
\right),\]

\begin{itemize}
\item photon (see (\ref{eq67})),
\end{itemize}

\[
\left. {U_{eff}^{s} } \right|_{r\to r_{0} } =-\dfrac{1}{\left( {r-r_{0} } 
\right)^{2}}\left( {\dfrac{1}{8}+\dfrac{r_{0}^{2} \omega^{2}}{2}} \right),
\]

\begin{itemize}
\item fermion with $\varepsilon =\varepsilon_{S}^{st} =0$ (see (\ref{eq84})),
\end{itemize}
\[
\left. {U_{eff}^{s} } \right|_{\rho \to 2\alpha } 
=-\dfrac{3}{32}\dfrac{1}{\left( {\rho -2\alpha } \right)^{2}}. 
\]

Reissner-Nordstr\"{o}m field:

\begin{itemize}
\item charged scalar particle and fermion with $\varepsilon \ne \varepsilon_{RN}^{st} $ (see 
(\ref{eq39})), 
\end{itemize}

\[
\left. {U_{eff} } \right|_{\rho \to \left( {\rho_{\pm } } \right)_{RN} } 
=-\,\,\dfrac{1}{\left( {\rho -\left( {\rho_{\pm } } \right)_{RN} } 
\right)^{2}}\left[ {\dfrac{1}{8}+\dfrac{\left( {\varepsilon -\alpha 
_{em} / \left( \rho_{\pm } \right)_{RN}} \right)^{2}\left( {\rho_{\pm 
} } \right)_{RN}^{4} }{2\left[ {\left( {\left( {\rho_{+} } \right)_{RN} 
-\left( {\rho_{-} } \right)_{RN} } \right)} \right]^{2}}} \right],
\]

\begin{itemize}
\item photon (see (\ref{eq67})),
\end{itemize}

\[
\left. {U_{eff} } \right|_{r\to \left( {r_{\pm } } \right)_{RN} } 
=-\,\,\dfrac{1}{\left( {r-\left( {r_{\pm } } \right)_{RN} } 
\right)^{2}}\left[ {\dfrac{1}{8}+\dfrac{\omega^{2}\left( {r_{\pm } } 
\right)_{RN}^{4} }{2\left[ {\left( {r_{+} } \right)_{RN} -\left( {r_{-} } 
\right)_{RN} } \right]^{2}}} \right],
\]

\begin{itemize}
\item fermion with $\varepsilon =\varepsilon_{RN}^{st} ={\alpha_{em} } \mathord{\left/ {\vphantom {{\alpha_{em} } {\left( {\rho_{\pm } } \right)_{RN} }}} \right. \kern-\nulldelimiterspace} {\left( {\rho_{\pm } } \right)_{RN} }$ (see (\ref{eq84})),
\end{itemize}

\[
\left. {U_{eff} } \right|_{\rho \to \left( {\rho_{\pm } } \right)_{RN} } 
=-\dfrac{3}{32}\dfrac{1}{\left( {\rho -\left( {\rho_{\pm } } \right)_{RN} } 
\right)^{2}}.
\]

Kerr, Kerr-Newman fields:

\begin{itemize}
\item uncharged scalar particle (see (\ref{eq39})),
\end{itemize}

\[
\left. {U_{eff}^{KN} } \right|_{\rho \to \left( {\rho_{\pm } } \right)_{KN} 
} =-\dfrac{1}{\left( {\rho -\left( {\rho_{\pm } } \right)_{KN} } 
\right)^{2}}\left[ {\dfrac{1}{8}+\dfrac{\left[ {\varepsilon \left( {\left( 
{\rho_{\pm } } \right)_{KN}^{2} +\alpha_{a}^{2} } \right)-\alpha_{a} 
m_{\varphi } } \right]^{2}}{2\left( {\left( {\rho_{+} } \right)_{KN} 
-\left( {\rho_{-} } \right)_{KN} } \right)^{2}}} \right],
\]

\begin{itemize}
\item photon (see 
(\ref{eq67})),
\end{itemize}

\[
\begin{array}{l}
\left. {U_{eff}^{KN} } \right|_{\rho \to \left( {\rho_{\pm } } \right)_{KN} 
} =-\,\,\dfrac{1}{\left( {r-\left( {r_{\pm } } \right)_{KN} } 
\right)^{2}} \times \\ [10pt]
\times \left[ {\dfrac{1}{8}+\dfrac{\left( {\omega -am_{\varphi } 
/ \left( r_{\pm }  \right)_{KN}^{2} +a^{2}} \right)^{2}\left( {\left( 
{r_{\pm } } \right)_{KN}^{2} +a^{2}} \right)^{2}}{2\left( {\left( {r_{+} } 
\right)_{KN} -\left( {r_{-} } \right)_{KN} } \right)^{2} }} \right], 
\end{array}
\]

\begin{itemize}
\item fermion with $\varepsilon \ne \varepsilon_{KN}^{st} $ (see (\ref{eq75})),
\end{itemize}

\[
\left. {U_{eff}^{KN} \left( {\varepsilon \ne \varepsilon_{KN}^{st} } 
\right)} \right|_{\rho \to \left( {\rho_{\pm } } \right)_{KN} } 
=-\dfrac{1}{\left( {\rho -\left( {\rho_{\pm } } \right)_{KN} } 
\right)^{2}}\left[ {\dfrac{1}{8}+\dfrac{\left( {\varepsilon -\varepsilon 
_{KN}^{st} } \right)^{2}\left( {\left( {\rho_{\pm } } \right)_{KN}^{2} 
+\alpha_{a}^{2} } \right)^{2}}{2\left[ {\left( {\rho_{+} } \right)_{KN} 
-\left( {\rho_{-} } \right)_{KN} } \right]^{2}}} \right],
\]

\begin{itemize}
\item fermion with $\varepsilon =\varepsilon_{KN}^{st} =\left[ \alpha_{a} m_{\varphi } +\alpha_{em} \left( \rho_{\pm } \right)_{KN} \right] / (\alpha_{a}^{2} +\left( {\rho_{\pm } } \right)_{KN}^{2} )  $ (see (\ref{eq84})),
\end{itemize}

\[
\left. {U_{eff}^{KN} } \right|_{\rho \to \left( {\rho_{\pm } } \right)_{KN} 
} =-\dfrac{3}{32}\dfrac{1}{\left( {\rho -\left( {\rho_{\pm } } \right)_{KN} } 
\right)^{2}}.
\]

Reissner-Nordstr\"{o}m extremal field:

\begin{itemize}
\item charged scalar particle with $\varepsilon \ne \varepsilon_{RN}^{ext} $, fermion with $\varepsilon \ne \varepsilon_{RN}^{ext} $ (see (\ref{eq49})),
\end{itemize}

\[
\left. {U_{RN}^{ext} } \right|_{\rho \to \alpha } =-\,\,\dfrac{\left( 
{\varepsilon -\varepsilon^{ext}} \right)^{2}\alpha^{4}}{2\left( {\rho 
-\alpha } \right)^{4}},
\]

\begin{itemize}
\item photon (see (\ref{eq69})),
\end{itemize}

\[
\left. {U_{RN}^{ext} } \right|_{r\to {r_{0} } \mathord{\left/ {\vphantom 
{{r_{0} } 2}} \right. \kern-\nulldelimiterspace} 2} =-\,\,\dfrac{\omega 
^{2}r_{0}^{4} }{2^{5}\left( {r-\left( {{r_{0} } \mathord{\left/ {\vphantom 
{{r_{0} } 2}} \right. \kern-\nulldelimiterspace} 2} \right)} \right)^{4}},
\]

\begin{itemize}
\item charged scalar particle with $\varepsilon =\varepsilon_{RN}^{ext} ={\alpha_{em} } \mathord{\left/ {\vphantom {{\alpha_{em} } \alpha }} \right. \kern-\nulldelimiterspace} \alpha $,
\end{itemize}

\[
\left. {U_{RN}^{ext} } \right|_{\rho \to \alpha } =-\,\,\dfrac{1}{2\left( 
{\rho -\alpha } \right)^{2}}\left[ {-l\left( {l+1} \right)-\alpha 
^{2}+\alpha_{em}^{2} } \right],
\]
\begin{itemize}
\item fermion with $\varepsilon =\varepsilon_{RN}^{ext} ={\alpha_{em} } \mathord{\left/ {\vphantom {{\alpha_{em} } \alpha }} \right. \kern-\nulldelimiterspace} \alpha $ (see (\ref{eq85}), (\ref{eq88})),
\end{itemize}

\[
\left. {U_{RN}^{ext} } \right|_{\rho \to \alpha } =-\,\,\dfrac{1}{2\left( 
{\rho -\alpha } \right)^{2}}\left[ {\dfrac{1}{4}-\kappa^{2}-\alpha 
^{2}+\alpha_{em}^{2} } \right].
\]

Kerr and Kerr-Newman extremal fields:

\begin{itemize}
\item uncharged scalar particle with $\varepsilon \ne \varepsilon^{ext}$, fermion with $\varepsilon \ne \varepsilon^{ext}$ (see (\ref{eq49})),
\end{itemize}

\[
\left. {U_{KN}^{ext} } \right|_{\rho \to \alpha } =-\,\,\dfrac{\left( {\alpha 
^{2}+\alpha_{a}^{2} } \right)^{2}\left( {\varepsilon -\varepsilon^{ext} 
} \right)^{2}}{2\left( {\rho -\alpha } \right)^{4}},
\]

\begin{itemize}
\item photon with $\omega \ne \omega^{ext}$ (see (\ref{eq69})),
\end{itemize}

\[
\left. {U_{KN}^{ext} } \right|_{r\to {r_{0} /2 } } =-\,\,\dfrac{\left( 
{r_{0}^{2} / 4 + \alpha^{2} } \right)^{2}\left( {\omega -\omega 
^{ext} } \right)^{2}}{2\left( {r-\left( {{r_{0} } \mathord{\left/ 
{\vphantom {{r_{0} } 2}} \right. \kern-\nulldelimiterspace} 2} \right)} 
\right)^{4}},
\]

\begin{itemize}
\item uncharged scalar particle with $\varepsilon =\varepsilon^{ext}=\alpha_{a} m_{\varphi } / (\alpha^{2} +\alpha_{a}^{2} )$ (see (\ref{eq50}), (\ref{eq51})),
\end{itemize}

\[
\left. {U_{KN}^{ext} } \right|_{\rho \to \alpha } =-\,\,\dfrac{1}{2\left( 
{\rho -\alpha } \right)^{2}}\left[ {\Omega^{2}-\alpha^{2}-\lambda 
_{ext} -C_{ext} } \right],
\]

\begin{itemize}
\item fermion with $\varepsilon =\varepsilon^{ext}=( \alpha_{a} m_{\varphi } +\alpha_{em} \alpha ) /(\alpha^{2}  +\alpha_{a}^{2} )$ (see (\ref{eq85})),
\end{itemize}

\[
\left. {U_{KN}^{ext} } \right|_{\rho \to \alpha } =-\,\,\dfrac{1}{2\left( 
{\rho -\alpha } \right)^{2}}\left[ {\dfrac{1}{4}-\left( {\lambda_{ext}^{2} 
+\alpha^{2}-\alpha^{4}\Omega^{2}} \right)} \right],
\]

\begin{itemize}
\item photon with $\omega =\omega^{ext}= m_{\varphi } \alpha / \left[  ( r_{0}^{2}/4)+\alpha^{2} \right] $ (see (\ref{eq70})),
\end{itemize}

\[
\left. {U_{KN}^{ext} } \right|_{r\to {r_{0} } \mathord{\left/ {\vphantom 
{{r_{0} } 2}} \right. \kern-\nulldelimiterspace} 2} =\,\,\dfrac{N_{ext} 
}{2\left( {r-\left( {{r_{0} } \mathord{\left/ {\vphantom {{r_{0} } 2}} 
\right. \kern-\nulldelimiterspace} 2} \right)} \right)^{2}}.
\]

For scalar particles, the final results can be supplemented with the Eddington-Finkelstein and Painlev\'{e} -Gullstrand metrics (see Sec. 3.2.3), for which the leading singularity in a neighborhood of the event horizon stays on only the real axis and is the same as for the Schwarzschild metric.

In addition, it was shown in \cite{43} for the Schwarzschild metric in isotropic coordinates and for the Eddington-Finkelstein, Painlev\'{e} -Gullstrand, Lemaitre-Finkelstein, and Kruskal-Szekeres metrics that in the case of a stationary bound state $\varepsilon^{st} =0$, leading singularity (\ref{eq84}) in a neighborhood of the event horizon has the same form as for the initial Schwarzschild metric (\ref{eq8}).

\section{Metrics with nonzero cosmological constant}
\label{sec:4}

\subsection{Kerr-Newman-(anti-)de Sitter geometry}
The stationary Kerr-Newman metric is characterized by a pointlike source with the mass $M$ and charge $Q$ rotating with the angular momentum $J=Mca$.

We can write the Kerr-Newman-(anti-)de Sitter metric in the Boyer-Lindquist coordinates $\left( 
{t,r,\theta ,\varphi } \right)$ in the form \cite{51} - \cite{55}.

\begin{equation}
\label{eq89}
\begin{array}{l}
 ds^{2}=\dfrac{\Delta_{r}^{KN} }{\Xi^{2}r_{K}^{2} }\left( {dt-a\sin 
^{2}\theta d\varphi } \right)^{2}-\dfrac{r_{K}^{2} }{\Delta_{r}^{KN} 
}dr^{2}-\dfrac{r_{K}^{2} }{\Delta_{\theta } }d\theta^{2}- \\ [10pt]
 -\dfrac{\Delta_{\theta } \sin^{2}\theta }{\Xi^{2}r_{K}^{2} }\left( 
{adt-\left( {r^{2}+a^{2}} \right)d\varphi } \right)^{2}, \\ 
 \end{array}
\end{equation}

\begin{equation}
\label{eq90}
\Delta_{\theta } =1+\frac{a^{2}\Lambda }{3}\cos^{2}\theta 
,\,\,\,\,\,\,\,\,\,\Xi =1+\frac{a^{2}\Lambda }{3},
\end{equation}

\begin{equation}
\label{eq91}
\Delta_{r}^{KN} =\left( {1-\frac{\Lambda }{3}r^{2}} \right)\left( 
{r^{2}+a^{2}} \right)-r_{0} r+r_{Q}^{2} ,
\end{equation}

\begin{equation}
\label{eq92}
r_{K}^{2} =r^{2}+a^{2}\cos^{2}\theta ,
\end{equation}
where $\Lambda $ is the cosmological constant, $r_{0} ={2GM} 
\mathord{\left/ {\vphantom {{2GM} {c^{2}}}} \right. 
\kern-\nulldelimiterspace} {c^{2}}$ is the gravitational radius, and  $r_{Q} 
={\sqrt G Q} \mathord{\left/ {\vphantom {{\sqrt G Q} {c^{2}}}} \right. 
\kern-\nulldelimiterspace} {c^{2}}$.

For $\Lambda >0$ (the de Sitter solution) in the presence of event horizons, we can represent $\Delta_{r}^{KN} $ given by (\ref{eq91}) in the form

\begin{equation}
\label{eq93}
\Delta_{r}^{KN} =-\frac{\Lambda }{3}\left( {r-r_{+} } \right)\left( 
{r-r_{-} } \right)\left( {r-r_{\Lambda }^{+} } \right)\left( {r-r_{\Lambda 
}^{-} } \right),
\end{equation}
where $r_{\pm } $ are the radii of the outer and inner event horizons and $r_{\Lambda }^{+} $ is the cosmological horizon.

For $\Lambda <0$ (the anti-de Sitter solution), the equation $\Delta 
_{r}^{KN} =0$ has two real and two complex-conjugate roots. We can represent $\Delta 
_{r}^{KN} $ in the form

\begin{equation}
\label{eq94}
\Delta_{r}^{KN} =\left( {r-r_{+} } \right)\left( {r-r_{-} } \right)\beta 
\left( r \right),
\end{equation}
where $\beta \left( r \right)$ is real function.

{\bf{4.1.1}} {\bf{Motion of scalar particles}}. For particles with  zero spin, the mass $m$, and the charge $q$, the Klein--Gordon-Fock equation in a curved space-time has a form

\begin{equation}
\label{eq95}
\left( {-g} \right)^{-1 \mathord{\left/ {\vphantom {1 2}} \right. 
\kern-\nulldelimiterspace} 2}\left( {\frac{\partial }{\partial x^{\mu 
}}-iqA_{\mu } } \right)\left[ {\left( {-g} \right)^{1 \mathord{\left/ 
{\vphantom {1 2}} \right. \kern-\nulldelimiterspace} 2}g^{\mu \nu }\left( 
{\frac{\partial }{\partial x^{\nu }}-iqA_{\nu } } \right)\Phi } 
\right]+m^{2}\Phi =0.
\end{equation}
For the Kerr-Newman-(anti-)de Sitter metric, Eq. (\ref{eq95}) admits separation of 
variables \cite{55}. If we set

\begin{equation}
\label{eq96}
\Phi \left( {{\rm {\bf r}},t} \right)=R\left( r \right)S\left( \theta 
\right)e^{im_{\varphi } \varphi }e^{-iEt},
\end{equation}
then we can write the equation for the radial function $R\left( r \right)$ in the form 

\begin{equation}
\label{eq97}
\frac{d^{2}R}{dr^{2}}+\frac{\left( {\Delta_{r}^{KN} } \right)^{\prime 
}}{\Delta_{r}^{KN} }\frac{dR}{dr}+\left[ {\Xi^{2}\left( {K-\frac{qQr}{\Xi 
}} \right)^{2}\frac{1}{\left( {\Delta_{r}^{KN} } 
\right)^{2}}-mr^{2}\frac{1}{\Delta_{r}^{KN} }-K_{lm_{\varphi } } 
\frac{1}{\Delta_{r}^{KN} }} \right]R=0,
\end{equation}
where $\left( {\Delta_{r}^{KN} } \right)^{\prime 
}=(d/dr) \Delta_{r}^{KN}$, $K=E\left( {r^{2}+a^{2}} 
\right)-am_{\varphi } $, $K_{lm_{\varphi } } $ is the separation constant, 
$E$ is the particle energy, $l=0,~1,~2,...$ is the quantum number of the 
particle orbital momentum, and  $m_{\varphi } =-l,-l+1,...l-1,l$ is orbital momentum projection .

We set

\begin{equation}
\label{eq98}
A=\dfrac{{\left( {\Delta_{r}^{KN} } \right)^{\prime }}}{\Delta_{r}^{KN} }, \,\,\,
B=\Xi^{2}\left( {K-\dfrac{qQr}{\Xi }} \right)^{2}\dfrac{1}{\left( {\Delta 
_{r}^{KN} } \right)^{2}}-mr^{2}\dfrac{1}{\Delta_{r}^{KN} }-K_{lm_{\varphi } 
} \dfrac{1}{\Delta_{r}^{KN} }.
\end{equation}
We bring Eq. (\ref{eq97}) to the form of the Schr\"{o}dinger equation with 
the effective potential $U_{eff} \left( r \right)$:

\begin{equation}
\label{eq100}
\bar{{R}}\left( r \right)=R\left( r \right)\exp \left[ \frac{1}{2}\int {A\left( 
{{r}'} \right)d} {r}'\right] ,
\end{equation}

\begin{equation}
\label{eq101}
\frac{d^{2}\bar{{R}}\left( r \right)}{dr^{2}}+2\left( {E_{Schr} -U_{eff} 
\left( r \right)} \right)\bar{{R}}\left( r \right)=0,
\end{equation}

\begin{equation}
\label{eq102}
U_{eff} \left( r \right)=E_{Schr} +\frac{1}{4} 
\frac{dA}{dr}+\frac{1}{8}A^{2}-\frac{1}{2}B,
\end{equation}

\begin{equation}
\label{eq103}
E_{Schr} =\frac{1}{2}\left( {E^{2}-m^{2}} \right).
\end{equation}
We can write effective potential (\ref{eq102}) in the explicit form

\begin{equation}
\label{eq104}
\begin{array}{l}
U_{eff} \left( r \right)=\dfrac{1}{2}\left( {E^{2}-m^{2}} 
\right)+\dfrac{1}{4}\dfrac{\left( {\Delta_{r}^{KN} } \right)^{\prime \prime 
}}{\Delta_{r}^{KN} }-\dfrac{1}{8}\dfrac{\left( {\Delta_{r}^{KN} \,^{\prime 
}} \right)^{2}}{\left( {\Delta_{r}^{KN} } \right)^{2}}- \\ [10pt]
-\Xi^{2}\left( 
{K-\dfrac{qQr}{\Xi }} \right)^{2}\dfrac{1}{2\left( {\Delta_{r}^{KN} } 
\right)^{2}}+\dfrac{mr^{2}}{2\Delta_{r}^{KN} }+\dfrac{K_{lm_{\varphi } } 
}{2\Delta_{r}^{KN} }.
\end{array}
\end{equation}
It can be seen from representations (\ref{eq93}) and (\ref{eq94}) that the leading singularities  $\sim 
1/\left( {r-r_{+} } \right)^{2}$, $1/\left( {r-r_{-} } \right)^{2}$, and $1/\left( {r-r_{\Lambda }^{+} } \right)^{2}$ near the horizons $r_{+} ,r_{-}$, and $r_{\Lambda }^{+} $ are contained in the third and forth terms in expression (\ref{eq104}).

The asymptotic formulas for effective potential (\ref{eq104}) have the same  structural form nearby horizons. For example, as $r\to r_{+} $,

\begin{equation}
\label{eq105}
\left. {U_{eff} } \right|_{r\to r_{+} } =-\frac{1}{\left( {r-r_{+} } 
\right)^{2}}\left\{ {\frac{1}{8}+\frac{\Xi^{2}\left( {K-qQr /\Xi } 
\right)^{2}}{2\left[ {\left( {r_{+} -r_{-} } \right)\left( {r_{+} 
-r_{\Lambda }^{+} } \right)\left( {r_{+} -r_{\Lambda }^{-} } \right)} 
\right]^{2}}} \right\}
\end{equation}
for the de Sitter solution $\left( {\Lambda >0} \right)$ and 

\begin{equation}
\label{eq106}
\left. {U_{eff} } \right|_{r\to r_{+} } =-\frac{1}{\left( {r-r_{+} } 
\right)^{2}}\left\{ {\frac{1}{8}+\frac{\Xi^{2}\left( {K-qQr / \Xi } 
\right)^{2}}{2\left[ {\left( {r_{+} -r_{-} } \right)\varphi \left( {r_{+} } 
\right)} \right]^{2}}} \right\}
\end{equation}
for the anti-de Sitter solution $\left( {\Lambda <0} \right)$. It can be seen from asymptotic formulas (\ref{eq105}) and  (\ref{eq106}) that for any scalar particle energy, near both sides of the horizons in potential (\ref{eq104}), there are infinitely deep potential wells \mbox{$\sim K_{1}^{+}/ \left( {r-r_{+} } 
\right)^{2}$, $K_{1}^{-} /\left( {r-r_{-} } 
\right)^{2}$}, and $K_{1}^{\Lambda }/ \left( {r-r_{\Lambda }^{+} } 
\right)^{2}$ with coefficients $K_{1}^{+,-,\Lambda } \ge 1/8$.

By criteria used in Sec. 2, the system ''a scalar 
particle in a Kerr-Newman-(anti-)de Sitter field'' is singular.

As horizons are approached, the radial function of a Schr\"{o}dinger-type 
equation has un unbounded number of zeros. For example, as $r\to r_{+} $,

\[
\left. {\bar{{R}}} \right|_{r\to r_{+} } \sim \left( {r-r_{+} } \right)^{1 
\mathord{\left/ {\vphantom {1 2}} \right. \kern-\nulldelimiterspace} 2}\sin 
\left( {\sqrt {L_{1}^{+} } \,\ln \left( {r-r_{+} } \right)+\delta } 
\right),
\]
where $\delta $  $\left( {0\le \delta <\pi } 
\right)$ is an arbitrary phase and $L_{1}^{+} =2\left( {K_{1}^{+} -\left( {1 \mathord{\left/ 
			{\vphantom {1 8}} \right. \kern-\nulldelimiterspace} 8} \right)} \right)$.

{\bf{4.1.2}} {\bf{Photon in a Kerr-Newman-(anti-)de Sitter field}}. Lunin \cite{48} separated the variables for the Kerr-(anti-)de Sitter metric in the form given in \cite{53}. 
The function $\psi $ in the master equation was presented in the form

\begin{equation}
\label{eq108}
\psi =e^{-i\omega \,t}e^{im\varphi }S\left( \theta \right)R\left( r 
\right).
\end{equation}
In the denotation in \cite{53} and with the signature of $ds^{2}$ in Sec. 4.1, we have (see (\ref{eq89}))

\begin{equation}
\label{eq109}
\begin{array}{l}
ds^{2}=-\tilde{{g}}_{tt} dt^{2}-\dfrac{r^{2}+a^{2}}{\Xi }\sin^{2}\theta 
\left[ {d\tilde{{\varphi }}-\dfrac{\Lambda }{3}adt} \right]^{2}-\\ [10pt]
-\dfrac{r_{0} 
r}{r_{K}^{2} }\left[ {dt-\dfrac{a\sin^{2}\theta d\tilde{{\varphi }}}{\Xi }} 
\right]^{2}-r_{K}^{2} \left[ {\dfrac{dr^{2}}{\Delta_{r}^{K} }+\dfrac{d\theta 
^{2}}{\Delta_{\theta } }} \right],
\end{array}
\end{equation}
where $\tilde{{g}}_{tt} =-\Delta_{\theta } \left( 1-\Lambda 
r^{2}/3 \right)/{\Xi }$ and $\Delta_{r}^{K} =\left( 1-\Lambda r^{2}/3 \right)\left( {r^{2}+a^{2}} \right)-r_{0} r$. The coordinate 
$\tilde{{\varphi }}$ has the standard periodicity $\left( {0\le 
\tilde{{\varphi }}<2\pi } \right)$, but to simplify some formulas, the angular coordinate

\begin{equation}
\label{eq110}
\varphi =\sqrt \Xi \,\tilde{{\varphi }},\,\,\,\,\,0\le \varphi <2\pi \sqrt \Xi,
\end{equation}
was introduced in \cite{48}. 
Just as in Sec. 3.3.2, we can separate the variables in the Maxwell equations 
in a Kerr-Newman-(anti-)de Sitter space-time using Lunin's results \cite{48} for the Kerr-(anti-)de Sitter geometry. For fermions, Chandrasekhar's paper \cite{49} was previously similarly generalized for the Kerr geometry by Page \cite{50}.

We must first change

\begin{equation}
\label{eq111}
\Delta^{K}\to \Delta_{r}^{KN} =\left( {1-\frac{\Lambda }{3}r^{2}} 
\right)\left( {r^{2}+a^{2}} \right)-r_{0} r+r_{Q}^{2}
\end{equation}
in the components of the Kinnersley tetrad. Further, following the Lunin formalism, we can separate the variables with ansatz (\ref{eq108}) and obtain formulas (2.72) in  \cite{48} for angular and radial equations with the charges $\Delta_{r}^{K} \to 
\Delta_{r}^{KN} $ and $r_{0} r\to \left( {r_{0} r-r_{Q}^{2} } \right)$ in them.

We emphasize that the ansatz (\ref{eq108}) contains the angle coordinate $\varphi $ 
with nonstandard periodicity (\ref{eq110}). Therefore, $m$ is not integer in the angular and radial 
equations, but  it takes discrete values as before.

We can write the radial equation for $R\left( r \right)$ in the form (see (2.72) in \cite{48})

\begin{equation}
\label{eq112}
\frac{d^{2}R}{dr^{2}}+\left( {\frac{\left( {\Delta_{r}^{KN} } 
\right)^{\prime }}{\Delta_{r}^{KN} }-\frac{{D}'_{r} }{D_{r} }} 
\right)\frac{dR}{dr}+\left[ {\frac{\left( {\omega \left( {r^{2}+a^{2}} 
\right)-am} \right)^{2}}{\left( {\Delta_{r}^{KN} } \right)^{2}}+\left( 
{\frac{2\xi }{D_{r} }-\xi } \right)\frac{1}{\Delta_{r}^{KN} }} 
\right]R=0,
\end{equation}
where $\left( {\Delta_{r}^{KN} } \right)^{\prime }=d\Delta 
_{r}^{KN} / dr ,\,\,\,{D}'_{r} =dD_{r} /dr$, $D_{r} =1+\left( {\mu r} \right)^{2}$  and $\xi =a\mu \left[ {m-a\omega 
	+\omega / a\mu^{2}} \right]$ for the "electric polarization",   $D_{r} =1+r^{2} /\left( {\mu a} \right)^{2}$ and  $\xi = (1 /\mu  )\left[ {a\omega -m-a\omega \mu^{2}} \right]$  for the "magnetic polarization",, and $\mu$ is the separation constant.

We set

\begin{equation}
\label{eq113}
A=\dfrac{{\left( {\Delta_{r}^{KN} } \right)^{\prime }}}{\Delta_{r}^{KN} },\,\,\,
B\left( r \right)=\frac{\left( {\omega \left( {r^{2}+a^{2}} \right)-am} 
\right)^{2}}{\left( {\Delta_{r}^{KN} } \right)^{2}}+\left( {\frac{2\xi 
}{D_{r} }-\xi } \right)\frac{1}{\Delta_{r}^{KN} }.
\end{equation}

We can then bring Eq. (\ref{eq112}) to the form of a Schr\"{o}dinger-type equation 
with the effective potential $U_{eff} \left( r \right)$ (see (\ref{eq100}) - 
(\ref{eq103})). Explicitly,

\begin{equation}
\label{eq115}
\begin{array}{l}
 U_{eff} =\dfrac{1}{2}\omega^{2}+\dfrac{1}{4}\left( {\dfrac{\left( {\Delta 
_{r}^{KN} } \right)^{\prime \prime }}{\Delta_{r}^{KN} }-\dfrac{{D}''_{r} 
}{D_{r} }} \right)+\dfrac{3}{8}\dfrac{{D}^{\prime\prime 2}_{r} }{D_{r}^{2} 
}-\dfrac{1}{4}\dfrac{\left( {\Delta_{r}^{KN} } \right)^{\prime }}{\Delta_{r}^{KN} }\dfrac{{D}^{\prime\prime}_{r} }{D_{r} }-\dfrac{1}{8}\dfrac{\left( {\Delta_{r}^{KN\prime}} \right)^{2}}{\left( {\Delta_{r}^{KN\,} } \right)^{2}}- \\ [5pt]
 -\dfrac{\left( {\omega \left( {r^{2}+a^{2}} \right)-am} \right)^{2}}{\left( 
{\Delta_{r}^{KN} } \right)^{2}}-\left( {\dfrac{2\xi }{D_{r} }-\xi } 
\right)\dfrac{1}{2\Delta_{r}^{KN} }. \\ 
 \end{array}
\end{equation}

It follows from representations (\ref{eq93}), (\ref{eq94}) that the leading singularities  $\sim 
1/\left( {r-r_{+} } \right)^{2}$, $1/ \left( {r-r_{-} } 
\right)^{2}$ and $1/ \left( {r-r_{\Lambda }^{+} } \right)^{2}$ near the horizons $r_{+} ,r_{-}$, and $r_{\Lambda }^{+} $ are contained in the fifth and sixth terms in expression (\ref{eq115}). Near the 
horizons, the asymptotic formulas for effective potential (\ref{eq115}) have the same structure. 
For example, as $r\to r_{+} $,

\begin{equation}
\label{eq116}
\left. {U_{eff} } \right|_{r\to r_{+} } =-\frac{1}{\left( {r-r_{+} } 
\right)^{2}}\left\{ {\frac{1}{8}+\frac{\left( {\omega \left( {r^{2}+a^{2}} 
\right)-am} \right)^{2}}{2\left[ {\left( {r_{+} -r_{-} } \right)\left( 
{r_{+} -r_{\Lambda }^{+} } \right)\left( {r_{+} -r_{\Lambda }^{-} } \right)} 
\right]^{2}}} \right\}
\end{equation}
for the de Sitter solution $\left( {\Lambda >0} \right)$ and

\begin{equation}
\label{eq117}
\left. {U_{eff} } \right|_{r\to r_{+} } =-\frac{1}{\left( {r-r_{+} } 
\right)^{2}}\left\{ {\frac{1}{8}+\frac{\left( {\omega \left( {r^{2}+a^{2}} 
\right)-am} \right)^{2}}{2\left[ {\left( {r_{+} -r_{-} } \right)\beta 
\left( {r_{+} } \right)} \right]^{2}}} \right\}
\end{equation}
for anti de Sitter solution $\left( {\Lambda <0} \right)$.

It follows from asymptotic formulas (\ref{eq116}) and (\ref{eq117}) that for any energy $\omega $, by both sides of the horizons in potential (\ref{eq115}), there are infinitely deep potential wells $\sim K_{2}^{+}/ \left( {r-r_{+} 
} \right)^{2}$, $K_{2}^{-} /{\left( {r-r_{-} } 
\right)^{2}}$, and $K_{2}^{\Lambda } / \left( {r-r_{\Lambda }^{+} } 
\right)^{2}$ with the coefficient $K_{2}^{+,-,\Lambda } \ge 1/8$. As a result, the regime of particle ''falling'' is realized on the corresponding event horizon, which is unacceptable in quantum theory.

As the horizons are approached, the radial function of the Schr\"{o}dinger-type 
equation has an unbounded number of zeros. For example, as $r\to r_{+} $,

\begin{equation}
\label{eq118}
\left. {\bar{{R}}\left( r \right)} \right|_{r\to r_{+} } \sim \left( 
{r-r_{+} } \right)^{1 \mathord{\left/ {\vphantom {1 2}} \right. 
\kern-\nulldelimiterspace} 2}\sin \left( {\sqrt {L_{2}^{+} } \,\ln \left( 
{r-r_{+} } \right)+\delta } \right),
\end{equation}
where $L_{2}^{+} =2\left( {K_{2}^{+} -\left( {1/ 8} \right)} \right)$.

In the Lunin formalism, the function $\psi $ in (\ref{eq108}) is related to the electromagnetic field potential $A^{\mu }\left( {{\rm {\bf r}},t} \right)$ by the components of the Kinnersley tetrad.

The components $A^{\mu }\left( {{\rm {\bf r}},t} \right)$ of the potential also behave as $\bar{{R}}\left( r 
\right)$ in (\ref{eq118}) in the neighborhood of horizons. Obviously, we can conclude that the system ''a photon in a Kerr-Newman-(anti-)de Sitter field'' is singular after quantization of the electromagnetic field.

{\bf{4.1.3}} {\bf{Fermion in a Kerr-Newman-(anti-)de Sitter field}}. Variables were separated in the Dirac equation in the Kerr-Newman-(anti-)de Sitter 
space-time in paper \cite{56}. Below, we use the form of the Kerr-Newman-(anti-)de Sitter metric and the notation in Sec. 4.1. (see (\ref{eq89}) - (\ref{eq94})).

In fact,  the Chandrasekhar ansatz \cite{49} 

\begin{equation}
\label{eq119}
\psi \left( {{\rm {\bf r}},t} \right)=\left( {{\begin{array}{*{20}c}
 {\dfrac{1}{r-ia\cos \theta }R^{\left( - \right)}\left( r \right)S^{\left( - 
\right)}\left( \theta \right)} \hfill \\ [5pt]
 {\dfrac{1}{\sqrt {\Delta_{r}^{KN} } }R^{\left( + \right)}\left( r 
\right)S^{\left( + \right)}\left( \theta \right)} \hfill \\[5pt]
 {\dfrac{1}{\sqrt {\Delta_{r}^{KN} } }R^{\left( + \right)}\left( r 
\right)S^{\left( - \right)}\left( \theta \right)} \hfill \\[5pt]
 {-\,\,\dfrac{1}{r+ia\cos \theta }R^{\left( - \right)}\left( r 
\right)S^{\left( + \right)}\left( \theta \right)} \hfill \\[5pt]
\end{array} }} \right)e^{-iEt+im_{\varphi } \varphi }
\end{equation}
was used to separate variables for the wave function 
$\psi $ of the Dirac equation with the mass $m$ and charge $q$ in \cite{56}, where the system of radial equations was obtained

\begin{equation}
\label{eq120}
\sqrt {\Delta_{r}^{KN} } \frac{dR^{\left( - \right)}}{dr}-\left[ 
{\frac{i\Xi \left( {E\left( {r^{2}+a^{2}} \right)-m_{\varphi } a} 
\right)-iqQr}{\sqrt {\Delta_{r}^{KN} } }} \right]R^{\left( - \right)}\left( 
r \right)=\left( {\lambda +imr} \right)R^{\left( + \right)}\left( r 
\right),
\end{equation}

\begin{equation}
\label{eq121}
\sqrt {\Delta_{r}^{KN} } \frac{dR^{\left( + \right)}}{dr}+\left[ 
{\frac{i\Xi \left( {E\left( {r^{2}+a^{2}} \right)-m_{\varphi } a} 
\right)-iqQr}{\sqrt {\Delta_{r}^{KN} } }} \right]R^{\left( + \right)}\left( 
r \right)=\left( {\lambda -imr} \right)R^{\left( - \right)}\left( r 
\right).
\end{equation}
Here $\lambda $ is separation constant, was obtained. It follows from (\ref{eq120}) and (\ref{eq121}) that $R^{\left( - \right)}\left( r \right)=R^{\left( + \right)\ast }\left( r 
\right)$. We introduce the real functions

\begin{equation}
\label{eq123}
\begin{array}{l}
g\left( r \right)=R^{\left( - \right)}\left( r \right)+R^{\left( + 
\right)}\left( r \right),\,\,\,
f\left( r \right)=-i\left( {R^{\left( - \right)}\left( r \right)-R^{\left( + 
\right)}\left( r \right)} \right).
\end{array}
\end{equation}
Adding (\ref{eq120}) and (\ref{eq121}) and subtracting (\ref{eq121}) from (\ref{eq120}), we obtain 

\begin{equation}
\label{eq125}
\begin{array}{l}
 \sqrt {\Delta_{r}^{KN} } \dfrac{df\left( r \right)}{dr}+\lambda f\left( r 
\right)-\left( {\dfrac{\Xi \left( {E\left( {r^{2}+a^{2}} \right)-m_{\varphi } 
a} \right)-qQr}{\sqrt {\Delta_{r}^{KN} } }+mr} \right)g\left( r \right)=0, 
\\ [5pt]
 \sqrt {\Delta_{r}^{KN} } \dfrac{dg\left( r \right)}{dr}-\lambda g\left( r 
\right)+\left( {\dfrac{\Xi \left( {E\left( {r^{2}+a^{2}} \right)-m_{\varphi } 
a} \right)-qQr}{\sqrt {\Delta_{r}^{KN} } }-mr} \right)f\left( r \right)=0. 
\\ 
 \end{array}
\end{equation}
We set $f_{KN} ={\Delta_{r}^{KN} }/{r^{2}}$ 
and introduce the functions $F_{KN} \left( r \right)=f\left( r 
\right) /r\sqrt {f_{KN} } $ and $G_{KN} \left( r \right)=g\left( r 
\right) / r\sqrt {f_{KN} }$. As a result, (\ref{eq125}) yields equations for the real radial functions $F_{KN} \left( r \right)$ and $G_{KN} \left( r \right)$:

\begin{equation}
\label{eq126}
\begin{array}{l}
 \left( {f_{KN} \dfrac{d}{dr}+\dfrac{1}{r}-\dfrac{r_{0} }{2r}+\dfrac{\lambda 
\sqrt {f_{KN} } }{r}} \right)F_{KN} \left( r \right)-\\ [10pt] 
-\left( {\Xi \left( 
{E\left( {1+\dfrac{a^{2}}{r^{2}}} \right)-\dfrac{m_{\varphi } a}{r^{2}}} 
\right)-\dfrac{qQ}{r}+mr\sqrt {f_{KN} } } \right)G_{KN} \left( r \right)=0, 
\\ [10pt]
 \left( {f_{KN} \dfrac{d}{dr}+\dfrac{1}{r}-\frac{r_{0} }{2r}-\dfrac{\lambda 
\sqrt {f_{KN} } }{r}} \right)G_{KN} \left( r \right)+\\ [10pt]
+\left( {\Xi \left( 
{E\left( {1+\dfrac{a^{2}}{r^{2}}} \right)-\dfrac{m_{\varphi } a}{r^{2}}} 
\right)-\dfrac{qQ}{r}-mr\sqrt {f_{KN} } } \right)F_{KN} \left( r \right)=0. 
\\ 
 \end{array}
\end{equation}

For $\Xi =1\,\,\,\left( {\Lambda =0} \right)$, Eqs. (\ref{eq126}) coincide 
with Eqs. (43) in \cite{6}, which were obtained in accordance with the results in \cite{57} using a more symmetric form of the Chandrasekhar-Page equations \cite{49}, \cite{50} and using Dirac matrices in the Dirac-Pauli representation.

Ansatz (\ref{eq119}) becomes

\begin{equation}
\label{eq127}
\psi \left( {{\rm {\bf r}},t} \right)=\frac{1}{\sqrt 2 }r\sqrt {f_{KN} } 
\left( {{\begin{array}{*{20}c}
 {F_{KN} \left( r \right)i\sigma^{3}\xi_{KN} \left( \theta \right)} 
\hfill \\
 {G_{KN} \left( r \right)\xi_{KN} \left( \theta \right)} \hfill \\
\end{array} }} \right)e^{-iEt+im_{\varphi } \varphi }
\end{equation}
with the spinor $\xi_{KN} \left( \theta \right)=\left( {{\begin{array}{*{20}c}
 {S_{-} \left( \theta \right)} \hfill \\
 {S_{+} \left( \theta \right)} \hfill \\
\end{array} }} \right)$, $S_{\pm } \left( \theta \right)$ are spheroidal 
harmonics for spin 1/2 satisfying the  Chandrasekhar-Page angular equations 
\cite{49}, \cite{50}.

Using the formalism in \cite{6} and taking the structural similarity of Eqs. (\ref{eq126}) and Eqs. (43) in 
\cite{6} into account, we can easily bring Eqs. (\ref{eq126}) to the 
form of a Schr\"{o}dinger-type equation with the effective potential $U_{eff} \left( \rho \right)$. The explicit form of the effective potential 
is given in Appendix B.

{\bf{4.1.3.1}} {\bf{Asymptotic formulas for the effective potential}}. 
It follows from representations (\ref{eq93}) and (\ref{eq94})  that effective potential (\ref{eq166})) (see Appendix B) near the horizons $r_{+},~r_{-}$, and $r_{\Lambda }^{+} $  is singular 
with the leading singularities of $\sim 1/ \left( {r-r_{+} } 
\right)^{2},\,\,1/ \left( {r-r_{-} } \right)^{2}$, and $1 / \left( {r-r_{\Lambda }^{+} } \right)^{2}$. The asymptotic formulas for the effective potential have the same 
structure near event horizons. For example, as $r\to r_{+} $,

\begin{equation}
\label{eq128}
\left. {U_{eff} } \right|_{r\to r_{+} } =-\frac{1}{\left( {r-r_{+} } 
\right)^{2}}\left\{ {\frac{1}{8}+\frac{\Omega_{+}^{2} }{2\left[ {\left( 
{r_{+} -r_{-} } \right)\left( {r_{+} -r_{\Lambda }^{+} } \right)\left( 
{r_{+} -r_{\Lambda }^{-} } \right)} \right]^{2}}} \right\}
\end{equation}
for the de Sitter solution $\left( {\Lambda >0} \right)$ and $\Omega_{+} \ne 0$ and

\begin{equation}
\label{eq129}
\left. {U_{eff} } \right|_{r\to r_{+} } =-\frac{1}{\left( {r-r_{+} } 
\right)^{2}}\left\{ {\frac{1}{8}+\frac{\Omega_{+}^{2} }{2\left[ {\left( 
{r_{+} -r_{-} } \right)\varphi \left( {r_{+} } \right)} \right]^{2}}} 
\right\}
\end{equation}
for the anti-de Sitter solution $\left( {\Lambda <0} \right)$ and $\Omega 
_{+} \ne 0$, where

\begin{equation}
\label{eq130}
\Omega_{+} =\Xi \left( {E\left( {r_{+}^{2} +a^{2}} \right)-m_{\varphi } 
a-qQr_{+} / \Xi } \right).
\end{equation}

The case $\Omega_{+} =0$ corresponds to a stationary state,

\begin{equation}
\label{eq131}
E_{+}^{st} =\frac{m_{\varphi } a+qQr_{+} / \Xi} {r_{+}^{2} +a^{2}}.
\end{equation}
This case is discussed in the next section.

Asymptotic formulas (\ref{eq128}) and (\ref{eq129}) show that for any energy $E\ne 
E^{st}$, there are infinitely deep potential wells 
$\sim K_{3}^{+} / \left( {r-r_{+} } \right)^{2}$, $K_{3}^{-} 
/ \left( {r-r_{-} } \right)^{2}$, and $K_{3}^{\Lambda } / \left( 
{r-r_{\Lambda }^{+} } \right)^{2}$ with the coefficients $K_{3}^{+,-,\Lambda } \ge 1/8$  in potential (\ref{eq166})). As a result, just as the preceding sections, the regime of particle ''falling'' on the corresponding event horizons is realized, which is unacceptable in quantum theory, and we conclude that for $E\ne E^{st}$, the system ''a fermion in a Kerr-Newman-(anti-)de Sitter'' field is singular.

{\bf{4.1.3.2} } {\bf{Fermion stationary states}}. We consider the case where either $\Omega_{+} =0$, $\Omega_{-} =0$, or $\Omega_{\Lambda }^{+} =0$. The quantities $\Omega_{-}$ and $\Omega_{\Lambda }^{+} $ have form (\ref{eq130}) with the change $r_{+} \to r_{-} 
$ or $r_{+} \to r_{\Lambda }^{+} $. In these cases, the fermion energy is equal to

\begin{equation}
\label{eq132}
E_{+,-,\Lambda^{+}}^{st} =\frac{m_{\varphi } a+qQr_{+,-,\Lambda^{+} 
}/ \Xi} {r_{+,-,\Lambda^{+}}^{2} +a^{2}}.
\end{equation}

In a neighborhood of event horizons, the asymptotic formula for the effective potential is (\ref{eq166})

\begin{equation}
\label{eq1333}
\left. {U_{eff} \left( {\Omega_{+,-,\Lambda^{+}} =0} \right)} 
\right|_{r\to r_{+,-,\Lambda^{+}} } =-\frac{3}{32}\frac{1}{\left( 
{r-r_{+,-,\Lambda^{+}} } \right)^{2}}.
\end{equation}

Expressions (\ref{eq128}) and (\ref{eq129}) do not coincide with asymptotic formula (\ref{eq1333}) as 
$\Omega_{+} \to 0$. For their coincidence, in expression (\ref{eq166}) for $U_{eff} $, terms that are insignificant for a finite $\Omega_{+} $ but noticeably contribute to the coefficient at the leading singularity as $\Omega_{+} \to 0$ must be taken into account. A similar remark also hold for expression (\ref{eq166}) for $U_{eff}$ as $\Omega_{-} \to 0$ or $\Omega_{\Lambda }^{+} \to 0$.

Asymptotic formula (\ref{eq1333}) for $\left| {E_{+,-,\Lambda^{+}}^{st} } \right|<m$ 
admits the existence of stationary bound states of spin-1/2 particles. Such states with a zero cosmological constant $\left( {\Xi =1} \right)$ were analyzed in \cite{4}--\cite{6}. Metrics with $\Xi \ne 1$ can be analyzed similarly. Solutions of  (\ref{eq132}) with $\left| {E_{+,-,\Lambda^{+}}^{st} } \right|<m$ correspond to a  Schr\"{o}dinger-type equation with square-integrable wave functions vanishing on event horizons. Particles in stationary states are located near event horizons with a high probability. Probability density maximums for detecting particles are separated from event 
horizons by fractions of the Compton wavelength of bound fermions.

\subsection{Other geometries}

Asymptotic formulas for effective potentials of a Schr\"{o}dinger-type equation for scalar particles, photons, and fermions were obtained for the most general Kerr-Newman-\mbox{(anti-)}de Sitter metric in Sec. 4.1. Analogous asymptotic formulas retain their structure for other geometries (Kerr-(anti-)de Sitter, Reissner-Nordstr\"{o}m-(anti-)de Sitter, and Schwarzschild-(anti-)de Sitter.

{\bf{4.2.1}} {\bf{Scalar particles}}. 
In asymptotic formulas (\ref{eq105}) and  (\ref{eq106}),

\begin{equation}
\label{eq134}
\begin{array}{l}
\Delta_{r}^{KN} =\left( {1-\frac{\Lambda }{3}} \right)\left( {r^{2}+a^{2}} 
\right)-r_{0} r+r_{Q}^{2} ,\,\,\,\,Q\ne 0,\\ [10pt]
K_{KN} =E\left( {r^{2}+a^{2}} 
\right)-am_{\varphi } ,\,\,\Xi =1+\frac{a^{2}\Lambda }{3},
\end{array}
\end{equation}
 for the Kerr-Newman metric,

\begin{equation}
\label{eq135}
\Delta_{r}^{K} =\left( {1-\frac{\Lambda }{3}} \right)\left( {r^{2}+a^{2}} 
\right)-r_{0} r,\,\,\,Q=0,\,\,\,K_{K} =E\left( {r^{2}+a^{2}} 
\right)-am_{\varphi } ,\,\,\Xi =1+\frac{a^{2}\Lambda }{3},
\end{equation}
for the Kerr metric,

\begin{equation}
\label{eq136}
\Delta_{r}^{RN} =\left( {1-\frac{\Lambda }{3}} \right)r^{2}-r_{0} 
r+r_{Q}^{2} ,\,\,\,Q\ne 0,\,\,\,K_{RN} =Er^{2},\,\,\Xi =1,
\end{equation}
for the Reissner-Nordstr\"{o}m metric, and 

\begin{equation}
\label{eq137}
\Delta_{r}^{S} =\left( {1-\frac{\Lambda }{3}} \right)r^{2}-r_{0} 
r,\,\,\,Q=0,\,\,\,r_{-} =0,\,\,K_{S} =Er^{2},\,\,\Xi =1,
\end{equation}
for the Schwarzschild metric.

{\bf{4.2.2}} {\bf{Photon}}. 
In asymptotic formulas (\ref{eq116}) and (\ref{eq117}), numerators in the second terms are 
equal to $K_{KN}^{2} $ with the change substitution $E\to \omega ,\,\,m_{\varphi } 
\to m$ (see (\ref{eq134})). It hence follows that we can pass to other metrics in accordance with formulas (\ref{eq135}) - (\ref{eq137}) with the change $E\to \omega ,\,\,m_{\varphi } \to m$.

{\bf{4.2.3}} {\bf{Fermion}}. 
The second terms in asymptotic formulas (\ref{eq128}) and  (\ref{eq129}) for fermions coincide with the second terms in asymptotic formulas (\ref{eq105}) and (\ref{eq106}) for scalar particles. It hence follows that we can pass to other metrics in accordance with (\ref{eq135}) - (\ref{eq137}).

{\bf{4.2.3.1}} {\bf{Stationary bound states of fermions}}. 
For the Kerr-Newman-(anti-)de Sitter metric, the energy of a fermion bound state is 
determined by expression (\ref{eq132}) with the condition

\begin{equation}
\label{eq138}
\left| {\frac{m_{\varphi } a+qQr_{+,-,\Lambda^{+}} / \Xi} 
{r_{+,-,\Lambda^{+}}^{2} +a^{2}}} \right|<m.
\end{equation}
Correspondingly, for the Kerr metric with $Q=0$, we have

\begin{equation}
\label{eq139}
\left( {E_{+,-,\Lambda^{+}}^{st} } \right)_{K} =\frac{m_{\varphi } 
a}{r_{+,-,\Lambda^{+}}^{2} +a^{2}}
\end{equation}
with the condition

\begin{equation}
\label{eq140}
\left| {\frac{m_{\varphi } a}{r_{+,-,\Lambda^{+}} +a^{2}}} \right|<m.
\end{equation}
For the Reissner-Nordstr\"{o}m metric with $a=0,\,\,Q\ne 0$, we have

\begin{equation}
\label{eq141}
\left( {E_{+,-,\Lambda^{+}}^{st} } \right)_{RN} =\frac{qQ}{\Xi 
r_{+,-,\Lambda^{+}} }
\end{equation}
with the condition 

\begin{equation}
\label{eq142}
\left| {\frac{qQ}{\Xi r_{+,-,\Lambda^{+}} }} \right|<m.
\end{equation}
For the Schwarzschild metric with $a=0,\,\,Q=0,\,\,r_{-} =0$, we have

\begin{equation}
\label{eq143}
\left( {E_{+,\Lambda^{+}}^{st} } \right)_{S} =0.
\end{equation}

\section{Kerr-Newman-anti-de Sitter five-dimensional geometry}

From physical standpoint, the five-dimensional anti-de Sitter black hole is interesting for using the Maldacena AdS/CFT correspondence. We represent the metric of a five-dimensional rotating charged Kerr-Newman-anti-de Sitter black hole in the Boyer-Lindquist coordinates $\left( {t,r,\theta ,\varphi ,\gamma } \right)$ with the Chern-Simons expression included in the form \cite{58}

\begin{equation}
\label{eq144}
\begin{array}{l}
 ds^{2}=g_{\mu \nu } dx^{\mu }dx^{\nu } =-\dfrac{\Delta_{r} }{\Sigma }X^{2}+\dfrac{\Sigma }{\Delta_{r} }dr^{2}+\dfrac{\Sigma }{\Delta_{\theta } }d\theta^{2}+ \\ [10pt]
+ \dfrac{\Delta_{\theta 
} \left( {a^{2}-b^{2}} \right)\sin^{2}\theta \cos^{2}\theta }{p^{2}\Sigma 
}Y^{2}+\left( {\dfrac{ab}{rp}Z+\dfrac{Qp}{r\Sigma }X} \right)^{2}, \\ 
 \end{array}
\end{equation}
and the gauge potential has the form

\begin{equation}
\label{eq145}
\mbox{A}=\frac{\sqrt 3 Q}{2\Sigma }X,
\end{equation}
where

\begin{equation}
\label{eq133}
\begin{array}{l}
X=dt-\dfrac{a\sin^{2}\theta }{\mbox{X}_{a} }d\varphi -\dfrac{b\cos^{2}\theta 
}{\mbox{X}_{b} }d\gamma , \\ [10pt]
Y=dt-\dfrac{\left( {r^{2}+a^{2}} \right)a}{\left( {a^{2}-b^{2}} 
\right)\mbox{X}_{a} }d\varphi -\dfrac{\left( {r^{2}+b^{2}} \right)a}{\left( 
{b^{2}-a^{2}} \right)\mbox{X}_{b} }d\gamma ,\\[10pt]
Z=dt-\dfrac{\left( {r^{2}+a^{2}} \right)\sin^{2}\theta }{a\mbox{X}_{a} 
}d\varphi -\dfrac{\left( {r^{2}+b^{2}} \right)\cos^{2}\theta }{b\mbox{X}_{b} 
}d\gamma ,\\ [10pt]
 \Delta_{r} =\left( {r^{2}+a^{2}} \right)\left( {r^{2}+b^{2}} \right)\left( 
{\dfrac{1}{r^{2}}+\dfrac{1}{l^{2}}} 
\right)-2M+\dfrac{Q^{2}+2Qab}{r^{2}},\,\,\,\Delta_{\theta } 
=1-\dfrac{p^{2}}{l^{2}}, \\  [10pt]
 \Sigma =r^{2}+p^{2},\,\,\,p=\sqrt {a^{2}\cos^{2}\theta +b^{2}\sin 
^{2}\theta } ,\,\,\,\mbox{X}_{a} =1-\dfrac{a^{2}}{l^{2}},\,\,\,\,\mbox{X}_{b} 
=1-\dfrac{b^{2}}{l^{2}}. \\ 
\end{array}
\end{equation}
Here, the parameters $\left( {M,Q,a,b,l} \right)$ depend on the mass, two 
independent black-hole angular momenta, and the cosmological constant.

Event horizons are defined by the equality $\Delta_{r} =0$. For 
example, the outer event horizon is determined by the largest root of the 
equation $\Delta_{r_{+} } =\left( {r-r_{+} } \right)\beta \left( 
{r_{+} } \right)=0$, where $\beta \left( {r_{+} } \right)\ne 0$.

\subsection{Motion of scalar particles}

In \cite{58}, variables were separated in the five-dimensional massive Klein-Gordon 
equation for a scalar field $\Phi \left( {t,r,\theta ,\varphi ,\gamma } 
\right)$ with metric (\ref{eq144}), (\ref{eq145}). With the variable separation ansatz $\Phi =R\left( r 
\right)S\left( p \right)e^{i\left( {m\,\varphi +k\gamma -\omega \,t} 
\right)}$, the equation for the radial function $R\left( r \right)$ has the form

\begin{equation}
\label{eq146}
\begin{array}{l}
 \dfrac{1}{r}\partial_{r} \left( {r\Delta_{r} \partial_{r} R} 
\right)+\left\{ {\dfrac{1}{r^{4}\Delta_{r} }} \right.\left[ {\left( 
{r^{2}+a^{2}} \right)} \right.\left( {r^{2}+b^{2}} \right)\omega -\left( 
{r^{2}+b^{2}} \right)ma\mbox{X}_{a} - \\ [10pt]
 \left. {-\left( {r^{2}+a^{2}} \right)kb\mbox{X}_{b} +Q\left( {ab\omega 
-mb\mbox{X}_{a} -ka\mbox{X}_{b} } \right)-\dfrac{\sqrt 3 }{2}qQr^{2}} 
\right]^{2}- \\ [10pt]
 \left. {-\dfrac{1}{r^{2}}\left( {ab\omega -mb\mbox{X}_{a} -ka\mbox{X}_{b} } 
\right)^{2}-\mu_{0}^{2} r^{2}-\lambda_{0}^{2} } \right\} R\left( r 
\right)=0, \\ 
 \end{array}
\end{equation}
where $\mu_{0}$ and $q$ are scalar particle mass and charge and $\lambda_{0} $ is the separation constant. We can bring Eq. (\ref{eq146}) to the form

\begin{equation}
\label{eq147}
\frac{d^{2}R}{dr^{2}}+A\frac{dR}{dr}+\left( {\frac{B_{1}^{2} }{\Delta 
_{r}^{2} }+\frac{B_{2} }{\Delta_{r} }} \right)R=0,
\end{equation}
where 

\begin{equation}
\label{eq148}
A=\frac{1}{r}+\frac{{\Delta }'_{r} }{\Delta_{r} },
\end{equation}

\begin{equation}
\label{eq149}
\begin{array}{l}
 B_{1} =\dfrac{1}{r^{2}}\left[ {\left( {r^{2}+a^{2}} \right)} \right.\left( 
{r^{2}+b^{2}} \right)\omega -\left( {r^{2}+b^{2}} \right)ma\mbox{X}_{a} - \\ [10pt]
 \left. {-\left( {r^{2}+a^{2}} \right)kb\mbox{X}_{b} +Q\left( {ab\omega 
-mb\mbox{X}_{a} -ka\mbox{X}_{b} } \right)-\dfrac{\sqrt 3 }{2}qQr^{2}} 
\right], \\ 
 \end{array}
\end{equation}

\begin{equation}
\label{eq150}
B_{2} =\frac{1}{r^{2}}\left( {ab\omega -mb\mbox{X}_{a} -ka\mbox{X}_{b} } 
\right)^{2}-\mu_{0}^{2} r^{2}-\lambda_{0}^{2} .
\end{equation}

Futher, we can bring Eq. (\ref{eq147}) to the form of a Schr\"{o}dinger equation with the effective potential $U_{eff} \left( r \right)$:

\begin{equation}
\label{eq151}
\bar{{R}}\left( r \right)=R\left( r \right)\exp \left[ \frac{1}{2}\int {A\left( 
{{r}'} \right)} \,d{r}'\right] ,
\end{equation}

\begin{equation}
\label{eq152}
\frac{d^{2}\bar{{R}}}{dr^{2}}+2\left( {E_{Schr} -U_{eff} \left( r \right)} 
\right)\bar{{R}}=0,
\end{equation}

\begin{equation}
\label{eq153}
U_{eff} \left( r \right)=E_{Schr} 
+\frac{1}{4}\frac{dA}{dr}+\frac{1}{8}A^{2}-\frac{1}{2}\frac{B_{1}^{2} 
}{\Delta_{r}^{2} }-\frac{B_{2} }{2\Delta_{r} },
\end{equation}

\begin{equation}
\label{eq154}
E_{Schr} =\frac{1}{2}\left( {E^{2}-m^{2}} \right).
\end{equation}
The term $E_{Schr} $ is distinguished in (\ref{eq152}) to give an equation of the 
Schr\"{o}dinger type. On the other hand, transferring this term into equality (\ref{eq153}) ensures the classical asymptotic form of the effective potential as $r\to \infty $.

We consider asymptotic formula (\ref{eq153}) in a neighborhood of the outer even horizon $r_{+} $. In this case, 

\begin{equation}
\label{eq155}
\Delta_{r} =\left( {r-r_{+} } \right)\beta \left( r 
\right),\,\,\,\,\left. {\beta \left( r \right)} 
\right|_{r\to r_{+} } \ne 0.
\end{equation}
The leading singularity of the expression 
$(1/4) (dA/dr)+(1/8) A^{2}$ in (\ref{eq153}) is equal to

\begin{equation}
\label{eq156}
\left. {\left( {\frac{1}{4}\frac{dA}{dr}+\frac{1}{8}A^{2}} \right)} 
\right|_{r\to r_{+} } =-\frac{1}{8\left( {r-r_{+} } \right)^{2}}.
\end{equation}
The leading singularity of the effective potential $U_{eff} \left( r 
\right)$ in a neighborhood of the outer event horizon is equal to

\begin{equation}
\label{eq157}
\left. {U_{eff} } \right|_{r\to r_{+} } =-\frac{1}{\left( {r-r_{+} } 
\right)^{2}}\left[ {\frac{1}{8}+\frac{B_{1}^{2} }{2\beta \left( {r_{+} } 
\right)^{2}}} \right].
\end{equation}
It can be seen from asymptotic formula (\ref{eq157}) that for any scalar particle energy, there are infinitely deep potential wells $\sim K_{+} / \left( {r-r_{+} } \right)^{2}$ with the 
coefficient $K_{+} \ge 1/8$  on both sides of the outer event horizon. In this case, the regime of a particle "falling" on the event horizon is realized, which is inconsistent with quantum mechanics, and by the criteria used in Sec. 2, the system ''a scalar particle in the field of a five-dimensional 
Kerr-Newman-anti-de Sitter black hole'' is singular.

As the event horizon $r_{+} $ is approached, the radial function of a Schr\"{o}dinger-type equation has an unbounded number of zeros,

\begin{equation}
\label{eq158}
\left. {\bar{{R}}} \right|_{r\to r_{+} } \sim \left( {r-r_{+} } \right)^{1 
\mathord{\left/ {\vphantom {1 2}} \right. \kern-\nulldelimiterspace} 2}\sin 
\left( {\sqrt {L_{+} } \ln \left( {r-r_{+} } \right)+\delta } \right),
\end{equation}
where $\delta $$\left( {0\le \delta \le \pi } 
\right)$ is arbitrary phase and $L_{+} =2\left( {K_{+} -\left( {1 \mathord{\left/ {\vphantom {1 
8}} \right. \kern-\nulldelimiterspace} 8} \right)} \right)$.

The inner event horizon $r_{-} $ can be considered similarly.

\subsection{Fermion in a five-dimensional Kerr-Newman-anti-de Sitter  field}
Variables were separated in the Dirac equation in the five-dimensional Kerr-Newman-anti-de Sitter  black hole space-time with a Chern-Simons expression in \cite{58} using the ansatz 

\begin{equation}
\label{eq159}
\sqrt {r+ip\gamma^{5}} \Psi =\left( {{\begin{array}{*{20}c}
 {R_{2} \left( r \right)S_{1} \left( p \right)} \hfill \\
 {R_{1} \left( r \right)S_{2} \left( p \right)} \hfill \\
 {R_{1} \left( r \right)S_{1} \left( p \right)} \hfill \\
 {R_{2} \left( r \right)S_{2} \left( p \right)} \hfill \\
\end{array} }} \right)e^{i\left( {m\varphi +k\gamma -\omega t} \right)}
\end{equation}
to separate the variables for the wave function $\Psi $ of the  five-dimensional Dirac 
equation with the fermion mass $\mu $ and the charge $q$. As a result, the system of equations for the radial functions $R_{1} \left( r \right), R_{2} \left( r \right)$ 

\begin{equation}
\label{eq160}
\sqrt {\Delta_{r} } \mbox{D}_{r}^{-} R_{1} =\left[ {\lambda +i\mu 
r-\frac{Q+ab}{2r^{2}}-\frac{i}{r}\left( {ab\omega -mb\mbox{X}_{a} 
-ka\mbox{X}_{b} } \right)} \right]R_{2},
\end{equation}

\begin{equation}
\label{eq161}
\sqrt {\Delta_{r} } \mbox{D}_{r}^{+} R_{2} =\left[ {\lambda -i\mu 
r-\frac{Q+ab}{2r^{2}}+\frac{i}{r}\left( {ab\omega -mb\mbox{X}_{a} 
-ka\mbox{X}_{b} } \right)} \right]R_{1} 
\end{equation}
was obtained, where  $\lambda $ is the separation constant and

\[
\begin{array}{l}
 \mbox{D}_{r}^{\pm } =\partial_{r} +\frac{{\Delta }'_{r} }{4\Delta_{r} 
}+\frac{1}{2r}\pm i\frac{1}{r^{2}\Delta_{r} }\left[ {\left( {r^{2}+a^{2}} 
\right)\left( {r^{2}+b^{2}} \right)\omega -\left( {r^{2}+b^{2}} 
\right)ma\mbox{X}_{a} -} \right. \\ [10pt]
 \left. {-\left( {r^{2}+a^{2}} \right)kb\mbox{X}_{b} +Q\left( {ab\omega 
-mb\mbox{X}_{a} -ka\mbox{X}_{b} } \right)-\frac{\sqrt 3 }{2}qQr^{2}} 
\right]. \\ 
 \end{array}
\]
It follows from the Eqs. (\ref{eq160}) and (\ref{eq161}) that

\begin{equation}
\label{eq162}
R_{1} \left( r \right)=R_{2}^{\ast } \left( r \right).
\end{equation}
We introduce the real functions

\begin{equation}
\label{eq163}
\begin{array}{l}
g\left( r \right)=R_{1} \left( r \right)+R_{2} \left( r \right),\,\,\,
f\left( r \right)=-i\left( {R_{1} \left( r \right)-R_{2} \left( r \right)} 
\right).
\end{array}
\end{equation}
Adding (\ref{eq160}) and (\ref{eq161}) and subtracting (\ref{eq161}) from (\ref{eq160}), we obtain

\begin{equation}
\label{eq165}
\begin{array}{l}
 \sqrt {\Delta_{r} } \dfrac{d}{dr}f+\left( {\dfrac{{\Delta }'_{r} }{4\sqrt 
{\Delta_{r} } }+\dfrac{\sqrt {\Delta_{r} } 
}{2r}-\dfrac{Q+ab}{2r^{2}}+\lambda } \right)f- \\ [10pt]
 -\dfrac{1}{r^{2}\sqrt {\Delta_{r} } }\left[ {\left( {r^{2}+a^{2}} 
\right)\left( {r^{2}+b^{2}} \right)\omega } \right.-\left( {r^{2}+b^{2}} 
\right)ma\mbox{X}_{a} -\left( {r^{2}+a^{2}} \right)kb\mbox{X}_{b} + \\ [10pt]
 \left. {+Q\left( {ab\omega -mb\mbox{X}_{a} -ka\mbox{X}_{b} } 
\right)-\dfrac{\sqrt 3 }{2}qQr^{2}} \right]g-\dfrac{1}{r}\left( {ab\omega 
-mb\mbox{X}_{a} -ka\mbox{X}_{b} } \right)g+\mu rg=0, \\ [10pt]
 \sqrt {\Delta_{r} } \dfrac{d}{dr}g+\left( {\dfrac{{\Delta }'_{r} }{4\sqrt 
{\Delta_{r} } }+\dfrac{\sqrt {\Delta_{r} } 
}{2r}+\frac{Q+ab}{2r^{2}}-\lambda } \right)g+ \\ [10pt]
 +\dfrac{1}{r^{2}\sqrt {\Delta_{r} } }\left[ {\left( {r^{2}+a^{2}} 
\right)\left( {r^{2}+b^{2}} \right)\omega } \right.-\left( {r^{2}+b^{2}} 
\right)ma\mbox{X}_{a} -\left( {r^{2}+a^{2}} \right)kb\mbox{X}_{b} + \\ [10pt]
 \left. {+Q\left( {ab\omega -mb\mbox{X}_{a} -ka\mbox{X}_{b} } 
\right)-\dfrac{\sqrt 3 }{2}qQr^{2}} \right]f+\dfrac{1}{r}\left( {ab\omega 
-mb\mbox{X}_{a} -ka\mbox{X}_{b} } \right)f-\mu rf=0. \\ 
 \end{array}
\end{equation}
We set $f_{r} ={\Delta_{r} } \mathord{\left/ {\vphantom 
{{\Delta_{r} } {r^{2}}}} \right. \kern-\nulldelimiterspace} {r^{2}}$ and introduce the functions $F\left( r \right)={f\left( r \right)} 
\mathord{\left/ {\vphantom {{f\left( r \right)} {r\sqrt {f_{r} } }}} \right. 
\kern-\nulldelimiterspace} {r\sqrt {f_{r} } }$ and $G\left( r \right)={g\left( 
r \right)} \mathord{\left/ {\vphantom {{g\left( r \right)} {r\sqrt {f_{r} } 
}}} \right. \kern-\nulldelimiterspace} {r\sqrt {f_{r} } }$. As a result, the 
equations for real radial functions $F\left( r \right)$ and $G\left( r 
\right)$ become

\begin{equation}
\label{eq167}
\begin{array}{l}
 \dfrac{dF}{dr}=A\left( r \right)F+B\left( r \right)G, \\ [10pt]
 \dfrac{dG}{dr}=C\left( r \right)F+D\left( r \right)G, \\ 
 \end{array}
\end{equation}
where

\begin{equation}
\label{eq1688}
\begin{array}{l}
A\left( r \right)=-\dfrac{1}{f_{r} }\left[ {\dfrac{2f_{r} 
}{r}+\dfrac{3}{4}{f}'_{r} +\left( {\dfrac{\lambda }{r}-\dfrac{Q+ab}{2r^{3}}} 
\right)\sqrt {f_{r} } } \right], \\ [10pt]
B\left( r \right)=\dfrac{1}{f_{r} }\left[ {\dfrac{B_{1} \left( r 
\right)}{r^{2}}+\left( {-B_{3} \left( r \right)+\mu } \right)\sqrt {f_{r} } 
} \right], \\ [10pt]
C\left( r \right)=-\dfrac{1}{f_{r} }\left[ {\dfrac{B_{1} \left( r 
\right)}{r^{2}}+\left( {B_{3} \left( r \right)-\mu } \right)\sqrt {f_{r} } } 
\right], \\ [10pt]
D\left( r \right)=-\dfrac{1}{f_{r} }\left[ {\dfrac{2f_{r} 
}{r}+\dfrac{3}{4}{f}'_{r} -\left( {\dfrac{\lambda }{r}-\dfrac{Q+ab}{2r^{3}}} 
\right)\sqrt {f_{r} } } \right], \\ [10pt]
B_{3} \left( r \right)=\dfrac{1}{r^{2}}\left( {ab\omega 
-mb\mbox{X}_{a} -ka\mbox{X}_{b} } \right),
\end{array}
\end{equation}
and the expression for $B_{1} \left( r \right)$ is given in (\ref{eq149}). Further, if we make the transformations

\begin{equation}
\label{eq172}
\begin{array}{c}
\psi_{F} =g_{F} F,\,\,\,\,\,
\psi_{G} =g_{G} G,\,\,\,\,\,
g_{F} =\exp \left( {\dfrac{1}{2}\int\limits^r {A_{F} \left( {{r}'} 
\right)d{r}'} } \right), \\
g_{G} =\exp \left( {\dfrac{1}{2}\int\limits^r {A_{G} \left( {{r}'} 
\right)d{r}'} } \right), \\
A_{F} \left( r \right)=-\dfrac{1}{B}\dfrac{dB}{dr}-A-D,\,\,\,\,\,
A_{G} \left( r \right)=-\dfrac{1}{C}\dfrac{dC}{dr}-A-D,
\end{array}
\end{equation}
then we obtain self-adjoint Schr\"{o}dinger-type equations for the functions $\psi_{F}$ and $\psi_{G} $ with the effective potential $U_{eff}^{F} \left( R \right)$ and $U_{eff}^{G} \left( R \right)$ :

\begin{equation}
\label{eq178}
\frac{d^{2}\psi_{F} }{dr^{2}}+2\left( {E_{Schr} -U_{eff}^{F} \left( r 
\right)} \right)\psi_{F} =0,
\end{equation}

\begin{equation}
\label{eq179}
\frac{d^{2}\psi_{G} }{dr^{2}}+2\left( {E_{Schr} -U_{eff}^{G} \left( r 
\right)} \right)\psi_{G} =0,
\end{equation}
where $E_{Schr} =\left( \omega^{2}-\mu^{2} \right)/2$. 
The equation for particles corresponds to Eq. (\ref{eq178}), and the equation for 
antiparticle corresponds to Eq. (\ref{eq179}).

For particles, the effective potential has the form

\begin{equation}
\label{eq181}
\begin{array}{l}
 U_{eff}^{F} \left( r \right)=E_{Schr} +\dfrac{3}{8}\left( 
{\dfrac{1}{B}\dfrac{dB}{dr}} 
\right)^{2}-\dfrac{1}{4}\dfrac{1}{B}\dfrac{d^{2}B}{dr^{2}}+ \\ [10pt]
+\dfrac{1}{4}\dfrac{d}{dr}\left( 
{A-D} \right)- 
 -\dfrac{1}{4}\dfrac{A-D}{B}\dfrac{dB}{dr}+\dfrac{1}{8}\left( {A-D} 
\right)^{2}+\dfrac{1}{2}BC. \\ 
 \end{array}
\end{equation}
Explicit expression (\ref{eq181}) has a cumbersome form. An expression 
for the effective potential in different geometries of four-dimensional space-time 
 was previously given many times in our papers \cite{4} - 
\cite{6}, where a more detailed presentation of the used formalism was also given.

{\bf{5.2.1}} {\bf{Asymptotic behavior of the effective potential}}. 
In the presence of outer and inner event horizons,

\[
\Delta_{r} =r^{2}f_{r} =\left( {r-r_{+} } \right)\left( {r-r_{-} } 
\right)\beta_{1} \left( r \right),
\]
formulas (\ref{eq1688}) show that effective potential (\ref{eq181}) is singular with leading singularities $\sim 1/ \left( {r-r_{+} } \right)^{2}$ and $\sim 1/ \left( {r-r_{-} } \right)^{2}$. The asymptotic formulas for the effective potential have the same structure near the event horizons. For example, in a 
neighborhood of the outer event horizon with  $B_{1} ( r_{+} ) \ne 0 $ as $r \to  $,

\begin{equation}
\label{eq182}
\left. {U_{eff} } \right|_{r\to r_{+} } =-\frac{1}{\left( {r-r_{+} } 
\right)^{2}}\left\{ {\frac{1}{8}+\frac{B_{1} \left( {r_{+} } 
\right)^{2}}{2\left[ {\left( {r_{+} -r_{-} } \right)\beta_{1} \left( 
{r_{+} } \right)} \right]^{2}}} \right\}.
\end{equation}
The case $B_{1} \left( {r_{+} } \right)=0$ corresponds to stationary states with the energy

\begin{equation}
\label{eq183}
\begin{array}{l}
 \omega_{+}^{st} =\dfrac{1}{\left( {r_{+}^{2} +a^{2}} \right)\left( 
{r_{+}^{2} +b^{2}} \right)+Qab}\left[ {\left( {r_{+}^{2} +b^{2}} 
\right)ma\mbox{X}_{a} +\left( {r_{+}^{2} +a^{2}} \right)kb\mbox{X}_{b} } 
\right.+ \\ [10pt]
 \left. {+Q\left( {mb\mbox{X}_{a} +ka\mbox{X}_{b} } \right)+\dfrac{\sqrt 3 
}{2}qQr_{+}^{2} } \right]. \\ 
 \end{array}
\end{equation}
This case is discussed in Sec. 5.2.2.

It follows from asymptotic formula (\ref{eq182}) that for any fermion energy $\omega \ne \omega^{st}$, there are infinitely deep potential wells $\sim K_{1}^{+}/ \left( r-r_{+}\right)^{2}$ and $\sim K_{1}^{-}/ \left( r-r_{-} \right)^{2}$ with coefficients $K_{1}^{\pm } \ge 1/ 8$ in potential (\ref{eq181}). 
As a results, just as for scalar particles, the regime of a fermion ''falling'' on the 
corresponding event horizons is realized, which is unacceptable in quantum theory. Therefore, we can conclude that for $\omega \ne \omega^{st}$, the system ''a fermion in a five-dimensional Kerr-Newman-anti-de Sitter field'' for is singular.

{\bf{5.2.2}} {\bf{Stationary  fermion states}}. 
We consider the case where either $B_{1} \left( {r_{+} } \right)=0$ or 
$B_{1} \left( {r_{-} } \right)=0$. In these cases, the fermion energy is equal to

\begin{equation}
\label{eq184}
\begin{array}{l}
 \omega_{\pm }^{st} =\dfrac{1}{\left( {r_{\pm }^{2} +a^{2}} \right)\left( 
{r_{\pm }^{2} +b^{2}} \right)+Qab}\left[ {\left( {r_{\pm }^{2} +b^{2}} 
\right)ma\mbox{X}_{a} +\left( {r_{\pm }^{2} +a^{2}} \right)kb\mbox{X}_{b} } 
\right.+ \\ [10pt]
 \left. {+Q\left( {mb\mbox{X}_{a} +ka\mbox{X}_{b} } \right)+\dfrac{\sqrt 3 
}{2}qQr_{\pm }^{2} } \right]. \\ 
 \end{array}
\end{equation}
Asymptotic formula  (\ref{eq181}) for the effective potential has the form 

\begin{equation}
\label{eq185}
\left. {U_{eff}^{F} } \right|_{r\to r_{\pm } } =-\frac{3}{32}\frac{1}{\left( 
{r-r_{\pm } } \right)^{2}}
\end{equation}
in a neighborhood of event horizon.

Expression (\ref{eq185}) does not coincide with asymptotic formula (\ref{eq182}) as $B_{1} 
\left( {r_{\pm } } \right)\to 0$. For coincidence, terms that are insignificant for a finite value $B_{1} $ but contribute noticebly to the coefficient for the leading singularity as $B_{1} 
\left( {r_{\pm } } \right)\to 0$ must be taken in account in the expression $U_{eff}^{F} $ (see (\ref{eq181})).

Asymptotic formula (\ref{eq185}) with $\left| {\omega_{\pm }^{st} } \right|<m$ admits 
the existence of stationary bound states of spin-1/2 particles. Such states were analyzed in \cite{4} - \cite{6} for different space-time geometries with a zero 
cosmological constant. Metrics with a nonzero cosmological constant, including the five-dimensional Kerr-Newman-anti-de Sitter black hole, can be analyzed similarly.

Solutions with $\left| {\omega_{\pm }^{st} } \right|<m$ correspond to 
square-integrable wave functions of a Schr\"{o}dinger-type equation 
vanishing on event horizons. Particles in stationary bound states are 
located near event horizons (above the outer event horizon and under 
the inner event horizon) with a high probability. The probability density maximums for detecting particles  are separated from the event horizons by fractions of the Compton wavelength of bound 
fermions.

\subsection{Photon in a five-dimensional Kerr field (Myers-Perry geometry) }

Lunin \cite{48} separated variables for the Maxwell equations in the five-dimensional 
Myers-Perry geometry. In that paper, there is also everything needed for separating variables for 
the Maxwell equations in the five-dimensional geometry of a rotating charged black hole with a nonzero cosmological constant. But we have here showed that the character of behavior of effective potential  in neighborhoods of event horizons is uncharged in passing from an uncharged to a charged rotating black hole. The same occurs in case of taking and not taking the cosmological constant into account in the metrics. Therefore, for brevity below, we restrict ourself to analyzing the behavior of the effective potential in a neighborhood of event horizons in the 
Myers-Perry geometry with an additional (fifth) dimension. In the 
notations in \cite{48}, the function $\Psi $ of the master equation is represented in the form

\begin{equation}
\label{eq186}
\Psi =e^{-i\omega t+im\,\varphi +in\,\gamma }\Phi \left( r \right)S\left( 
\theta \right).
\end{equation}

Ansatz (\ref{eq186}) leads to two types of solutions, which Lunin called 
''electrical'' and ''magnetic'' polarizations. The leading singularities of 
the effective potentials are the same for the two solution types. Below, as an example, we consider the ''electrical'' solution.

The equation for the radial function $\Phi \left( r \right)$ was given in 
formulas (4.31) in \cite{48}. Futher, we can 
obtain a Schr\"{o}dinger-type equation with an effective potential for the function $\bar{{\Phi }}\left( r 
\right)$ in the standard way (see Sec. 3.2). In Lunin's notation, the 
leading singularities of the effective potential near the event horizons $r_{\pm } $ have the form

\begin{equation}
\label{eq187}
\begin{array}{l}
\left. {U_{eff} } \right|_{r\to r_{\pm } } =-\dfrac{1}{\left( {r-r_{\pm } } 
\right)^{2}}\times \\[10pt]
\times \left\{ {\dfrac{1}{8}+\dfrac{Mr^{2}}{2R}\dfrac{\left[ {\left( 
{r^{2} +a^{2}} \right)\left( {r^{2} +b^{2}} \right)\omega -\left( 
{r^{2} +b^{2}} \right)ma-\left( {r^{2} +a^{2}} \right)nb} 
\right]^{2}}{\left[ {\left( {r_{\pm } -r_{\mp } } \right)\beta_{1} \left( 
{r_{\pm } } \right)} \right]^{2}}} \right\},
\end{array}
\end{equation}
where $R=\left( {r^{2} +a^{2}} \right)\left( {r^{2} +b^{2}} 
\right)$ and $\Delta_{r} =\left( {r-r_{+} } \right)\left( {r-r_{-} } 
\right)\beta_{1} \left( r \right)$.

It follows from asymptotic formula (\ref{eq187}) that for any energy $\omega $ in the effective potential on both sides of 
the event horizons $r_{\pm } $, there are infinitely deep potential wells $\sim {K_{2}^{\pm } } / {\left( {r-r_{\pm } } \right)^{2}}$ with 
coefficients $K_{2}^{\pm } \ge 1/8$. As a result, the regime of a particle ''falling'' 
on event horizons is realized, which is unacceptable in quantum theory.

As the event horizon is approached, the radial function $\bar{{\Phi 
}}\left( r \right)$ of a Schr\"{o}dinger-type equation has an unbounded 
number of zeros. For example, as $r\to r_{+} $,

\begin{equation}
\label{eq188}
\left. {\bar{{\Phi }}\left( r \right)} \right|_{r\to r_{+} } \sim \left( 
{r-r_{+} } \right)^{1 \mathord{\left/ {\vphantom {1 2}} \right. 
\kern-\nulldelimiterspace} 2}\sin \left( {\sqrt {L_{2}^{+} } \ln \left( 
{r-r_{+} } \right)+\delta } \right),
\end{equation}
where $L_{+} =2\left( {K_{2}^{+} -\left( {1 /8} \right)} \right)$.

In the Lunin formalism \cite{48}, the function $\Psi $ 
in (\ref{eq186}) is related to electromagnetic field potentials
$A^{\mu }\left( {t,r,\theta ,\varphi ,\gamma } \right)$ by the components of the generalized Kinnersley tetrad.

In a neighborhood of the event horizons, the oscillating behavior of 
$\bar{{\Phi }}\left( r \right)$ in (\ref{eq188}) is also present for the components 
$A^{\mu }\left( {t,r,\theta ,\varphi ,\gamma } \right)$. Obviously, after quantization of the electromagnetic field, we can conclude that the system ''a photon in a five-dimensional Myers-Perry field'' is singular.

\section{Discussion of results}
\label{sec:6}

Our analysis shows that for $\varepsilon \ne \varepsilon^{st}$ and $\varepsilon \ne \varepsilon^{ext} $ in the space-time of the considered black holes, the existence of stationary states of quantum particles is impossible. States of the systems ''a particle in fields of classical black holes with event horizons of zero thickness'' are singular. The existence of stationary discrete states with $\varepsilon 
=\varepsilon^{st}$ and $\varepsilon =\varepsilon^{ext} $ does not 
change the preceding conclusion, because to attain the values $\varepsilon 
^{st}$ and $\varepsilon^{ext} $, quantum transitions with emission or absorption of photons with a particular energy are necessary, but quantum mechanical stationary states of photons with the real energy $\omega $ do not exist in the considered gravitational and electromagnetic fields.

The universal character of divergence of the effective potentials near the event 
horizons is typical for all considered metrics and for particles with 
different spins. The discovered singularities do not allow applying quantum theory in full, which leads to the necessity to change the formulation of the original physical problem.

As a result of our research, two questions arise.

1. Can the solutions of general relativity that are quantum mechanically "ill-behaved" be cured?

The answer to this question seems to be negative. Indeed, the uniqueness theorem for 
black holes \cite{59} states that the most general asymptotically flat vacuum solution of the equations of general relativity theory is the Kerr metric with a monopole mass $M$ and the angular momentum $J$. Any deviation from a spherically symmetric mass distribution leads to the event horizon vanishing
and the occurence of several naked singularities in its place (see static and stationary $q$-metrics in \cite{60} - \cite{62}).

If by analogy with the Coulomb potential for $Z\ge 137\,\,\,\left( {\kappa 
=-1} \right)$, we match the outer vacuum solutions of general relativity to inner solution variants with preserving the continuity of the metric tensor and its first derivatives, then the event horizon vanishes, and the matching radius as a rule turns to be greater than the event horizon radius (see, e. g., \cite{63} - \cite{66}). Hence, in the considered cases, we pass beyond the concept of classical black holes with event horizons.

2. Can the solutions of general relativity that are quantum mechanically "ill-behaved" be used?

We answer this question positively and propose to supplement the 
gravitational collapse mechanism.

In the final stage of collapse, let the gravitational field capture spin-1/2 particles that after the formation of event horizons are in stationary bound states with $\varepsilon =\varepsilon^{st} $ both under the inner and above the outer event horizons. For the subsequent fermions interacting with such composite systems, the self-consistent gravitational and 
electromagnetic fields are determined both by the collapsar mass and charge and by the masses and charges of the fermions in stationary bound states with $\varepsilon 
=\varepsilon^{st} $ located near the event horizons. Obviously, such a 
system can be nonsingular. For a rigorous proof, we need precise calculations of the self-consistent gravitational and electromagnetic fields of the composite systems and a proof of the existence of stationary states of quantum mechanical test particles in them. The discussed composite systems can be building blocks for combining new particles and finally forming macroscopic objects. On the other hand, these 
systems can be regarded as carriers of dark matter \cite{4}, \cite{5}.

\section{Conclusion}
\label{sec:7}

For all considered metrics of the classical black holes and for particles spins, we have established the existence of the quantum mechanical regime of particle ''falling'' on event horizons.

We usde the Schwarzschild coordinates for the Schwarzschild and Reissner-Nordstr\"{o}m metrics and the Boyer-Lindquist coordinates for the Kerr and Kerr-Newman metrics. The transformation from the Schwarzschild coordinates to Eddington-Finkelstein and Painlev\'{e} -Gullstrand coordinates deso not eliminate the problem of a particle ''falling''.

The nonstationary Lema\^{i}tre -Finkelstein and Kruskal-Szekeres metrics lead to Hamiltonians depending on the time coordinates \cite{43}. In these cases, it is impossible to study stationary states with a representation of the wave functions in the coordinates of these metrics in form (\ref{eq1}), namely, in the form  $\sim \psi ({\bf R}) e^{-iET}$ for the Lema\^{i}tre -Finkelstein metric and in the form $\sim \psi ({\bf u}) e^{-iE\upsilon}$ for the Kruskal-Szekeres metric.

\section*{APPENDIX A}

\subsection*{Effective potential of a Painlev\'{e}-Gullstrand field in a Schr\"{o}dinger-type equation for a scalar particle}

We have

\[
\begin{array}{l}
	U_{eff}^{PG} \left( \rho \right)=-\dfrac{\alpha \left( {\varepsilon^{2}-1} 
		\right)}{\rho -2\alpha }-\dfrac{\alpha \rho \varepsilon^{2}}{\left( {\rho 
			-2\alpha } \right)^{2}}+ \dfrac{1}{2\left( {\rho -2\alpha } \right)^{2}}\left( 
	{\dfrac{\alpha^{2}}{\rho^{2}}+\dfrac{\alpha }{\rho }-1} 
	\right)+ \\ [10pt]
	+\dfrac{1}{2\rho \left( {\rho -2\alpha } \right)}\left( 
	{1+\dfrac{\alpha }{\rho }} \right)- \dfrac{1}{2\rho \left( {\rho -2\alpha } \right)}l\left( {l+1} 
	\right)+i\dfrac{1}{4}\sqrt {\dfrac{2\alpha }{\rho }} \dfrac{\varepsilon }{\rho 
	}, \\ 
\end{array}
\]
\[
\left. {U_{eff}^{PG} } \right|_{\rho \to \infty } =\frac{\alpha }{\rho 
}\left( {1-2\varepsilon^{2}} \right),\,\,\,\, \left. {U_{eff}^{PG} } \right|_{\rho \to 0} =-\frac{1}{8\rho^{2}},
\]
\[
\left. {U_{eff}^{PG} } \right|_{\rho \to 2\alpha } =-\,\,\frac{1}{2\left( 
	{\rho -2\alpha } \right)^{2}}\left( {\frac{1}{4}+4\alpha^{2}\varepsilon 
	^{2}} \right).
\]

\section*{APPENDIX B}

\subsection*{Effective potentials of gravitational and electromagnetic fields in Schr\"{o}dinger-type equations for fermions}

\begin{enumerate}
	\item For the Kerr-Newman-(anti-)de Sitter field, in accordance with \cite{6} and Eqs. (\ref{eq126}),
	
	\begin{equation}
		\label{eq166}
		\begin{array}{l}
			U_{eff}^{KN} =E_{Schr} +\dfrac{3}{8}\dfrac{1}{B_{KN}^{2} }\left( 
			{\dfrac{dB_{KN} }{dr}} \right)^{2}-\dfrac{1}{4B_{KN} 
			}\dfrac{d^{2}B_{KN} }{dr^{2}}+\dfrac{1}{4}\dfrac{d}{dr}\left( {A_{KN} 
				-D_{KN} } \right)- \\ [10pt]
			-\dfrac{1}{4}\dfrac{\left( {A_{KN} -D_{KN} } \right)}{B_{KN} 
			}\dfrac{dB_{KN} }{dr}+\dfrac{1}{8}\left( {A_{KN} -D_{KN} } 
			\right)^{2}+\dfrac{1}{2}B_{KN} C_{KN} , \\ 
		\end{array}
	\end{equation}
	
	\begin{equation}
		\label{eq167}
		\begin{array}{l}
			\dfrac{3}{8}\dfrac{1}{B_{KN}^{2} }\left( {\dfrac{dB_{KN} }{dr}} 
			\right)^{2}= \\[10pt]
			= \dfrac{3}{8}\left\{ {\dfrac{f_{KN} }{\Omega_{KN} +\sqrt {f_{KN} } 
				}\left[ {-\dfrac{1}{f_{KN}^{2} }{f}'_{KN} \left( {\Omega_{KN} +\sqrt {f_{KN} 
					} } \right)+\dfrac{1}{f_{KN} }\left( {{\Omega }'_{KN} +\dfrac{{f}'_{KN} 
						}{2\sqrt {f_{KN} } }} \right)} \right]} \right\}^{2},
		\end{array}
	\end{equation}
	
	\begin{equation}
		\label{eq168}
		\begin{array}{l}
			-\dfrac{1}{4}\dfrac{1}{B_{KN} }\dfrac{d^{2}B_{KN} 
			}{dr^{2}}=-\dfrac{1}{4}\dfrac{f_{KN} }{\Omega_{KN} +\sqrt {f_{KN} } }\left[ 
			{\dfrac{2}{f_{KN}^{3} }\left( {{f}'_{KN} } \right)^{2}\left( {\Omega_{KN} 
					+\sqrt {f_{KN} } } \right)-} \right. \\ [10pt]
			\left. {-\dfrac{1}{f_{KN}^{2} }{f}''_{KN} \left( {\Omega_{KN} +\sqrt 
					{f_{KN} } } \right)-\dfrac{2}{f_{KN}^{2} }{f}'_{KN} \left( {{\Omega }'_{KN} 
					+\dfrac{{f}'_{KN} }{2\sqrt {f_{KN} } }} \right)+} \right. \\
			\left. {+\dfrac{1}{f_{KN} }\left( 
				{{\Omega }''_{KN} +\dfrac{{f}''_{KN} }{2\sqrt {f_{KN} } }-\dfrac{\left( 
						{{f}'_{KN} } \right)^{2}}{4f_{KN}^{3 \mathord{\left/ {\vphantom {3 2}} 
								\right. \kern-\nulldelimiterspace} 2} }} \right)} \right], \\ 
		\end{array}
	\end{equation}
	
	\begin{equation}
		\label{eq169}
		\frac{1}{4}\frac{d}{dr}\left( {A-D} \right)=\frac{\lambda }{2}\left[ 
		{\frac{1}{2}\frac{{f}'_{KN} }{rf_{KN}^{3 \mathord{\left/ {\vphantom {3 2}} 
						\right. \kern-\nulldelimiterspace} 2} }+\frac{1}{r^{2}f_{KN}^{1 
					\mathord{\left/ {\vphantom {1 2}} \right. \kern-\nulldelimiterspace} 2} }} 
		\right],
	\end{equation}
	
	\begin{equation}
		\label{eq170}
		-\frac{1}{4}\frac{\left( {A-D} \right)}{B}\frac{dB}{dr}=\frac{\lambda 
		}{2rf_{KN}^{1 \mathord{\left/ {\vphantom {1 2}} \right. 
					\kern-\nulldelimiterspace} 2} }\left( {-\frac{{f}'_{KN} }{f_{KN} 
			}+\frac{1}{\Omega_{KN} +\sqrt {f_{KN} } }\left( {{\Omega }'_{KN} 
				+\frac{{f}'_{KN} }{2\sqrt {f_{KN} } }} \right)} \right),
	\end{equation}
	
	\begin{equation}
		\label{eq171}
		\begin{array}{l}
			\dfrac{1}{8}\left( {A-D} \right)^{2}=\dfrac{\lambda^{2}}{2f_{KN} r^{2}},\,\,\,
			\dfrac{1}{2}BC=-\dfrac{1}{2f_{KN}^{2} }\left( {\Omega_{KN}^{2} -f_{KN} } 
			\right),
		\end{array}
	\end{equation}
	where 
	
	\[
	\begin{array}{l}
		f_{KN} =\left( {1-\dfrac{\Lambda }{3}r^{2}} \right)\left( 
		{1+\dfrac{a^{2}}{r^{2}}} \right)-\dfrac{r_{0} }{r}+\dfrac{r_{Q}^{2} }{r^{2}},\\[10pt] 
		{f}'_{KN} \equiv \dfrac{df_{KN} }{dr}=-\dfrac{2\Lambda 
		}{3}r-\dfrac{2\alpha^{2} }{3r^{3}}+\dfrac{r_{0} }{r^{2}}-\dfrac{2r_{Q}^{2} 
		}{r^{3}}, \\ [10pt]
		{f}''_{KN} \equiv \dfrac{d^{2}f_{KN} 
		}{dr^{2}}=-\dfrac{2\Lambda }{3}-\dfrac{2r_{0} }{r^{3}}+\dfrac{2\alpha^{2} 
			+6r_{Q}^{2} }{r^{4}}, \\ [10pt]
		\Omega_{KN} =\Xi \left[ {E\left( {1+\dfrac{\alpha 
					^{2} }{r^{2}}} \right)-\dfrac{\alpha m_{\varphi } }{r^{2}}-\dfrac{qQ}{\Xi 
				r}} \right], \\ [10pt]
		{\Omega }'_{KN} \equiv \frac{d\Omega_{KN} }{dr}=\Xi \left[ 
		{-\dfrac{2E\alpha^{2} }{r^{3}}+\dfrac{2\alpha m_{\varphi } 
			}{r^{3}}+\dfrac{qQ}{\Xi r^{2}}} \right], \\ [10pt]
		{\Omega }''_{KN} \equiv 
		\dfrac{d^{2}\Omega_{KN} }{dr^{2}}=\Xi \left[ {\dfrac{6E\alpha^{2} 
			}{r^{4}}-\dfrac{6\alpha m_{\varphi } }{r^{4}}-\dfrac{2qQ}{\Xi r^{3}}} 
		\right].
	\end{array}
	\]
	The arithmetic sum of the expressions $E_{Schr} =\left( E^{2}-m^{2}\right)/2$ and relations (\ref{eq167}) - (\ref{eq171}) results in an expression for the effective 
	potential $U_{eff}^{F} $. For the remaining electromagnetic and gravitational 
	fields considered here, the structure of the expressions for the effective 
	potentials is unchanged. Only the expressions for 
	$f,\,\,{f}',\,\,{f}'',\,\,\Omega ,\,\,{\Omega }'$, and ${\Omega }''$ change.

	\item For the Kerr-(anti-)de Sitter field $\left( {Q=0\,} \right)$,
	
	\[
	\begin{array}{l}
		f_{K} =\left( {1-\dfrac{\Lambda }{3}r^{2}} \right)\left( 
		{1+\dfrac{a^{2}}{r^{2}}} \right)-\dfrac{r_{0} }{r},\,\,\,\,
		{f}'_{K} =-\dfrac{2\Lambda }{3}r-\dfrac{2\alpha^{2} 
		}{3r^{3}}+\dfrac{r_{0} }{r^{2}}, \\ [10pt]
		\quad
		{f}''_{K} =-\dfrac{2\Lambda }{3}-\dfrac{2r_{0} }{r^{3}}+\dfrac{2a^{2} 
		}{r^{4}},\\ [10pt]
		\Omega_{K} =\Xi \left[ {E\left( {1+\dfrac{\alpha^{2} }{r^{2}}} 
			\right)-\dfrac{\alpha m_{\varphi } }{r^{2}}} \right],\,\,\,\,
		{\Omega }'_{K} =\Xi \left[ {-\dfrac{2E\alpha^{2} }{r^{3}}+\dfrac{2\alpha 
				m_{\varphi } }{r^{3}}} \right],\,\,\,\,
		{\Omega }''_{K} =\Xi \left[ {\dfrac{6E\alpha^{2} }{r^{4}}-\dfrac{6\alpha 
				m_{\varphi } }{r^{4}}} \right]\,.
	\end{array}
	\]
	
	\item For the Reissner-Nordstr\"{o}m-(anti-)de Sitter field $\left( {a=0} \right)$,
	
	\[
	\begin{array}{l}
		f_{RN} =1-\dfrac{\Lambda }{3}r^{2}-\dfrac{r_{0} }{r}+\dfrac{r_{Q}^{2} 
		}{r^{2}},\,\,\,\,
		{f}'_{RN} =-\dfrac{2\Lambda }{3}r+\dfrac{r_{0} }{r^{2}}-\dfrac{2r_{Q}^{2} 
		}{r^{3}},\\ [10pt]
		{f}''_{RN} =-\dfrac{2\Lambda }{3}-\dfrac{2r_{0} 
		}{r^{3}}+\dfrac{6r_{Q}^{2} }{r^{4}},\\ [10pt]
		\Omega_{RN} =\Xi \,E-\dfrac{qQ}{r},\,\,\,\,
		{\Omega }'_{RN} =\frac{qQ}{r^{2}},\,\,\,\,\,
		{\Omega }''_{RN} =-\frac{2qQ}{r^{3}},\,\,\,\lambda =\kappa ,
	\end{array}
	\]
	where $\kappa $ is the separation constant,
	
	\[
	\kappa =\mp 
	1,\mp 2...=\left\{ {\begin{array}{l}
			-\left( {l-1} \right),\,\,\,j=l+\dfrac{1}{2}\\ 
			\,\,\,\,\,\,\,\,\,l,\,\,\,\,\,\,\,\,\,\,\,\,\,\,\,j=l-\dfrac{1}{2}  
	\end{array}} \right.,
	\]
and $j$ and $l$ are the quantum numbers of the total and orbital momenta of a spin-1/2 particle.
	
	\item For the Schwarzschild-(anti-)de Sitter field $\left( {Q=0,\,\,a=0} \right)$,
	
	\[
	\begin{array}{l}
		f_{S} =1-\dfrac{\Lambda }{3}r^{2}-\dfrac{r_{0} }{r},
		\quad
		{f}'_{S} =-\dfrac{2\Lambda }{3}r+\dfrac{r_{0} }{r^{2}},
		\quad
		{f}''_{S} =-\dfrac{2\Lambda }{3}-\dfrac{2r_{0} }{r^{3}},\\ [10pt]
		\Omega_{S} =\Xi \,E,\,\,\,\,
		{\Omega }'_{S} ={\Omega }''_{S} =0,\,\,\,\,\lambda =\kappa .
	\end{array}
	\]
\end{enumerate}

\subsection*{Acknowledgments}

The authors thank E. Yu. Popov for the useful discussions and the help in establishing the final form of some analytical expressions. The authors also thank A.L.Novoselova for the essential 
technical support in preparation the paper.

\subsection*{Conflicts of Interest}

The authors declare no conflicts of interest.



\end{document}